# The IceCube Neutrino Observatory VI:

# Neutrino Oscillations, Supernova Searches, Ice Properties

THE ICECUBE COLLABORATION

# IceCube Collaboration Member List

R. Abbasi[28], Y. Abdou[22], T. Abu-Zayyad[33], M. Ackermann[39], J. Adams[16], J. A. Aguilar[28], M. Ahlers[32], M. M. Allen[36], D. Altmann[1], K. Andeen[28,a], J. Auffenberg[38], X. Bai[31,b], M. Baker[28], S. W. Barwick[24], R. Bay[7], J. L. Bazo Alba[39], K. Beattie[8], J. J. Beatty[18,19], S. Bechet[13], J. K. Becker[10], K.-H. Becker[38], M. L. Benabderrahmane[39], S. BenZvi[28], J. Berdermann[39], P. Berghaus[31], D. Berley[17], E. Bernardini[39], D. Bertrand[13], D. Z. Besson[26], D. Bindig[38], M. Bissok[1], E. Blaufuss[17], J. Blumenthal[1], D. J. Boersma[1], C. Bohm[34], D. Bose[14], S. Böser[11], O. Botner[37], A. M. Brown[16], S. Buitink[14], K. S. Caballero-Mora[36], M. Carson[22], D. Chirkin[28], B. Christy[17], F. Clevermann[20], S. Cohen[25], C. Colnard[23], D. F. Cowen[36,35], A. H. Cruz Silva[39], M. V. D'Agostino[7], M. Danninger[34], J. Daughhetee[5], J. C. Davis[18], C. De Clercq[14], T. Degner[11], L. Demirörs[25], F. Descamps[22], P. Desiati[28], G. de Vries-Uiterweerd[22], T. DeYoung[36], J. C. Díaz-Vélez[28], M. Dierckxsens[13], J. Dreyer[10], J. P. Dumm[28], M. Dunkman[36], J. Eisch[28], R. W. Ellsworth[17], O. Engdegård[37], S. Euler[1], P. A. Evenson[31], O. Fadiran[28], A. R. Fazely[6], A. Fedynitch[10], J. Feintzeig[28], T. Feusels[22], K. Filimonov[7], C. Finley[34], T. Fischer-Wasels[38], B. D. Fox[36], A. Franckowiak[11], R. Franke[39], T. K. Gaisser[31], J. Gallagher[27], L. Gerhardt[8,7], L. Gladstone[28], T. Glüsenkamp[39], A. Goldschmidt[8], J. A. Goodman[17], D. Góra[39], D. Grant[21], T. Griesel[29], A. Gross[16,23], S. Grullon[28], M. Gurtner[38], C. Ha[36], A. Haj Ismail[22], A. Hallgren[37], F. Halzen[28], K. Han[39], K. Hanson[13,28], D. Heinen[1], K. Helbing[38], R. Hellauer[17], S. Hickford[16], G. C. Hill[28], K. D. Hoffman[17], A. Homeier[11], K. Hoshina[28], W. Huelsnitz[17,c], J.-P. Hülss[1], P. O. Hulth[34], K. Hultqvist[34], S. Hussain[31], A. Ishihara[15], E. Jacobi[39], J. Jacobsen[28], G. S. Japaridze[4], H. Johansson[34], K.-H. Kampert[38], A. Kappes[9], T. Karg[38], A. Karle[28], P. Kenny[26], J. Kiryluk[8,7], F. Kislat[39], S. R. Klein[8,7], J.-H. Köhne[20], G. Kohnen[30], H. Kolanoski[9], L. Köpke[29], S. Kopper[38], D. J. Koskinen[36], M. Kowalski[11], T. Kowarik[29], M. Krasberg[28], T. Krings[1], G. Kroll[29], N. Kurahashi[28], T. Kuwabara[31], M. Labare[14], K. Laihem[1], H. Landsman[28], M. J. Larson[36], R. Lauer[39], J. Lünemann[29], J. Madsen[33], A. Marotta[13], R. Maruyama[28], K. Mase[15], H. S. Matis[8], K. Meagher[17], M. Merck[28], P. Mészáros[35,36], T. Meures[13], S. Miarecki[8,7], E. Middell[39], N. Milke[20], J. Miller[37], T. Montaruli[28,d], R. Morse[28], S. M. Movit[35], R. Nahnhauer[39], J. W. Nam[24], U. Naumann[38], D. R. Nygren[8], S. Odrowski[23], A. Olivas[17], M. Olivo[10], A. O'Murchadha[28], S. Panknin[11], L. Paul[1], C. Pérez de los Heros[37], J. Petrovic[13], A. Piegsa[29], D. Pieloth[20], R. Porrata[7], J. Posselt[38], P. B. Price[7], G. T. Przybylski[8], K. Rawlins[3], P. Redl[17], E. Resconi[23,e], W. Rhode[20], M. Ribordy[25], M. Richman[17], J. P. Rodrigues[28], F. Rothmaier[29], C. Rott[18], T. Ruhe[20], D. Rutledge[36], B. Ruzybayev[31], D. Ryckbosch[22], H.-G. Sander[29], M. Santander[28], S. Sarkar[32], K. Schatto[29], T. Schmidt[17], A. Schönwald[39], A. Schukraft[1], A. Schultes[38], O. Schulz[23,f], M. Schunck[1], D. Seckel[31], B. Semburg[38], S. H. Seo[34], Y. Sestayo[23], S. Seunarine[12], A. Silvestri[24], G. M. Spiczak[33], C. Spiering[39], M. Stamatikos[18,g], T. Stanev[31], T. Stezelberger[8], R. G. Stokstad[8], A. Stössl[39], E. A. Strahler[14], R. Ström[37], M. Stüer[11], G. W. Sullivan[17], Q. Swillens[13], H. Taavola[37], I. Taboada[5], A. Tamburro[33], A. Tepe[5], S. Ter-Antonyan[6], S. Tilav[31], P. A. Toale[2], S. Toscano[28], D. Tosi[39], N. van Eijndhoven[14], J. Vandenbroucke[7], A. Van Overloop[22], J. van Santen[28], M. Vehring[1], M. Voge[11], C. Walck[34], T. Waldenmaier[9], M. Wallraff[1], M. Walter[39], Ch. Weaver[28], C. Wendt[28], S. Westerhoff[28], N. Whitehorn[28], K. Wiebe[29], C. H. Wiebusch[1], D. R. Williams[2], R. Wischnewski[39], H. Wissing[17], M. Wolf[23], T. R. Wood[21], K. Woschnagg[7], C. Xu[31], D. L. Xu[2], X. W. Xu[6], J. P. Yanez[39], G. Yodh[24], S. Yoshida[15], P. Zarzhitsky[2], M. Zoll[34]

[1]III. Physikalisches Institut, RWTH Aachen University, D-52056 Aachen, Germany
[2]Dept. of Physics and Astronomy, University of Alabama, Tuscaloosa, AL 35487, USA
[3]Dept. of Physics and Astronomy, University of Alaska Anchorage, 3211 Providence Dr., Anchorage, AK 99508, USA
[4]CTSPS, Clark-Atlanta University, Atlanta, GA 30314, USA
[5]School of Physics and Center for Relativistic Astrophysics, Georgia Institute of Technology, Atlanta, GA 30332, USA
[6]Dept. of Physics, Southern University, Baton Rouge, LA 70813, USA
[7]Dept. of Physics, University of California, Berkeley, CA 94720, USA
[8]Lawrence Berkeley National Laboratory, Berkeley, CA 94720, USA
[9]Institut für Physik, Humboldt-Universität zu Berlin, D-12489 Berlin, Germany
[10]Fakultät für Physik & Astronomie, Ruhr-Universität Bochum, D-44780 Bochum, Germany
[11]Physikalisches Institut, Universität Bonn, Nussallee 12, D-53115 Bonn, Germany
[12]Dept. of Physics, University of the West Indies, Cave Hill Campus, Bridgetown BB11000, Barbados
[13]Université Libre de Bruxelles, Science Faculty CP230, B-1050 Brussels, Belgium
[14]Vrije Universiteit Brussel, Dienst ELEM, B-1050 Brussels, Belgium
[15]Dept. of Physics, Chiba University, Chiba 263-8522, Japan
[16]Dept. of Physics and Astronomy, University of Canterbury, Private Bag 4800, Christchurch, New Zealand
[17]Dept. of Physics, University of Maryland, College Park, MD 20742, USA
[18]Dept. of Physics and Center for Cosmology and Astro-Particle Physics, Ohio State University, Columbus, OH 43210, USA
[19]Dept. of Astronomy, Ohio State University, Columbus, OH 43210, USA
[20]Dept. of Physics, TU Dortmund University, D-44221 Dortmund, Germany
[21]Dept. of Physics, University of Alberta, Edmonton, Alberta, Canada T6G 2G7
[22]Dept. of Physics and Astronomy, University of Gent, B-9000 Gent, Belgium
[23]Max-Planck-Institut für Kernphysik, D-69177 Heidelberg, Germany
[24]Dept. of Physics and Astronomy, University of California, Irvine, CA 92697, USA
[25]Laboratory for High Energy Physics, École Polytechnique Fédérale, CH-1015 Lausanne, Switzerland
[26]Dept. of Physics and Astronomy, University of Kansas, Lawrence, KS 66045, USA
[27]Dept. of Astronomy, University of Wisconsin, Madison, WI 53706, USA
[28]Dept. of Physics, University of Wisconsin, Madison, WI 53706, USA
[29]Institute of Physics, University of Mainz, Staudinger Weg 7, D-55099 Mainz, Germany
[30]Université de Mons, 7000 Mons, Belgium
[31]Bartol Research Institute and Department of Physics and Astronomy, University of Delaware, Newark, DE 19716, USA
[32]Dept. of Physics, University of Oxford, 1 Keble Road, Oxford OX1 3NP, UK
[33]Dept. of Physics, University of Wisconsin, River Falls, WI 54022, USA
[34]Oskar Klein Centre and Dept. of Physics, Stockholm University, SE-10691 Stockholm, Sweden
[35]Dept. of Astronomy and Astrophysics, Pennsylvania State University, University Park, PA 16802, USA
[36]Dept. of Physics, Pennsylvania State University, University Park, PA 16802, USA
[37]Dept. of Physics and Astronomy, Uppsala University, Box 516, S-75120 Uppsala, Sweden
[38]Dept. of Physics, University of Wuppertal, D-42119 Wuppertal, Germany
[39]DESY, D-15735 Zeuthen, Germany
[a]now at Dept. of Physics and Astronomy, Rutgers University, Piscataway, NJ 08854, USA
[b]now at Physics Department, South Dakota School of Mines and Technology, Rapid City, SD 57701, USA
[c]Los Alamos National Laboratory, Los Alamos, NM 87545, USA
[d]also Sezione INFN, Dipartimento di Fisica, I-70126, Bari, Italy
[e]now at T.U. Munich, 85748 Garching & Friedrich-Alexander Universität Erlangen-Nürnberg, 91058 Erlangen, Germany
[f]now at T.U. Munich, 85748 Garching, Germany
[g]NASA Goddard Space Flight Center, Greenbelt, MD 20771, USA

## Acknowledgments

We acknowledge the support from the following agencies: U.S. National Science Foundation-Office of Polar Programs, U.S. National Science Foundation-Physics Division, University of Wisconsin Alumni Research Foundation, the Grid Laboratory Of Wisconsin (GLOW) grid infrastructure at the University of Wisconsin - Madison, the Open Science Grid (OSG) grid infrastructure; U.S. Department of Energy, and National Energy Research Scientific Computing Center, the Louisiana Optical Network Initiative (LONI) grid computing resources; National Science and Engineering Research Council of Canada; Swedish Research Council, Swedish Polar Research Secretariat, Swedish National Infrastructure for Computing (SNIC), and Knut and Alice Wallenberg Foundation, Sweden; German Ministry for Education and Research (BMBF), Deutsche Forschungsgemeinschaft (DFG), Research Department of Plasmas with Complex Interactions (Bochum), Germany; Fund for Scientific Research (FNRS-FWO), FWO Odysseus programme, Flanders Institute to encourage scientific and technological research in industry (IWT), Belgian Federal Science Policy Office (Belspo); University of Oxford, United Kingdom; Marsden Fund, New Zealand; Japan Society for Promotion of Science (JSPS); the Swiss National Science Foundation (SNSF), Switzerland; D. Boersma acknowledges support by the EU Marie Curie IRG Program; A. Groß acknowledges support by the EU Marie Curie OIF Program; J. P. Rodrigues acknowledges support by the Capes Foundation, Ministry of Education of Brazil; A. Schukraft acknowledges the support by the German Telekom Foundation; N. Whitehorn acknowledges support by the NSF Graduate Research Fellowships Program.

# Atmospheric neutrino oscillations with DeepCore

THE ICECUBE COLLABORATION[1]
[1]*See special section in these proceedings*

**Abstract:** IceCube DeepCore can study atmospheric neutrino oscillations through a combination of its low energy reach, as low as about 10 GeV, and its unprecedented statistical sample, of about 150,000 triggered atmospheric muon neutrinos per year. With the diameter of the earth as a baseline, the muon neutrino disappearance minimum and tau neutrino appearance maximum are expected at about 25 GeV, which is considerably lower energy than typical IceCube neutrino events, but higher than the energies at which accelerator-based experiments have detected oscillations. $\nu_\mu$ disappearance and $\nu_\tau$ appearance from neutrino oscillation can be measured in IceCube DeepCore. We present here the status of the newly developed low energy reconstruction algorithms, the expected experimental signatures, and the proposed approach for such neutrino oscillation measurements.

**Corresponding authors:** Sebastian Euler[2] (seuler@icecube.wisc.edu), Laura Gladstone[3](gladstone@icecube.wisc.edu), Jason Koskinen[4](koskinen@psu.edu), Donglian Xu[5](dxu@crimson.ua.edu)

[2]*III. Physikalisches Institut, RWTH Aachen University, D-52056 Aachen, Germany*
[3]*Dept. of Physics, University of Wisconsin, Madison, WI 53706, USA*
[4]*Dept. of Physics, Pennsylvania State University, University Park, PA 16802, USA*
[5]*Dept. of Physics and Astronomy, University of Alabama, Tuscaloosa, AL 35487, USA*

## 1 IceCube DeepCore

IceCube is a cubic-kilometer neutrino observatory located at the geographic South Pole. IceCube construction began in 2004 and was completed in December 2010. The complete detector consists of 86 strings deployed into the glacial ice, each of which consists of 60 Digital Optical Modules (DOMs) located between depths of 1450 m and 2450 m. Seventy-eight strings are arranged on a hexagonal grid with an average 125 m horizontal spacing and 17 m vertical DOM spacing. The remaining 8 strings are more closely spaced in the center of the detector, with horizontal distances of 40 - 70 m and vertical DOM spacing of 7 m. The 8 inner densely instrumented strings, optimized for low energies, together with the surrounding 12 IceCube standard strings, form the DeepCore inner detector (Fig. 1). These 8 inner strings have 10 DOMs located between 1750 m and 1850 m in depth and 50 DOMs located between 2100 m and 2450 m. The ice at depths between 1970 m and 2100 m, formed about 65,000 years ago [1], has a relatively short absorption length, and is known as the "dust layer". The 10 DOMs on each DeepCore string deployed above the dust layer help reject cosmic ray muons which are the major background to atmospheric neutrino studies. The ice below 2100 m has a scattering length about twice that of the ice in the upper part of the IceCube detector [2]. The lower 50 DeepCore DOMs are deployed in this very clear ice. DeepCore DOMs contain high quantum efficiency photomultipliers (HQE PMTs [3]) which add $\sim$ 35% increase in efficiency compared to the standard IceCube DOMs. With denser string and DOM spacing, clearer ice, as well as higher efficiency PMTs, DeepCore is optimized for low energy neutrino physics [4]. Fig. 2 shows the predicted $\nu_\mu$ and $\nu_e$ effective areas at both trigger and online veto levels. Below 100 GeV the addition of DeepCore increases the effective area of IceCube by more than an order of magnitude.

## 2 Neutrino Oscillation Physics in DeepCore

The IceCube DeepCore sub-array has opened a new window on atmospheric neutrino oscillation physics with its low energy reach to about 10 GeV. The oscillation measurement is also made feasible by DeepCore's location at the bottom center of IceCube, which allows the surrounding IceCube strings to act as an active veto against cosmic ray muons, the primary background to atmospheric neutrino measurements. A muon background rejection of $8 \times 10^{-3}$ for the overall IceCube trigger ($\sim$ 2000 Hz) is achieved by applying a veto algorithm which rejects events with particle speed (defined as the speed of a particle traveling from the hit in the surrounding region to the center of gravity (COG) in DeepCore) between 0.25 m/ns and 0.4 m/ns. The scheme of this veto algorithm is demonstrated in Fig. 3. With further expected improvements in the veto and event reconstruction algorithms, DeepCore expects to achieve a cosmic-ray muon rejection factor of $10^6$ or better [6]. Specific methods to investigate the oscillation phenomenon will be discussed in detail in the following subsections.

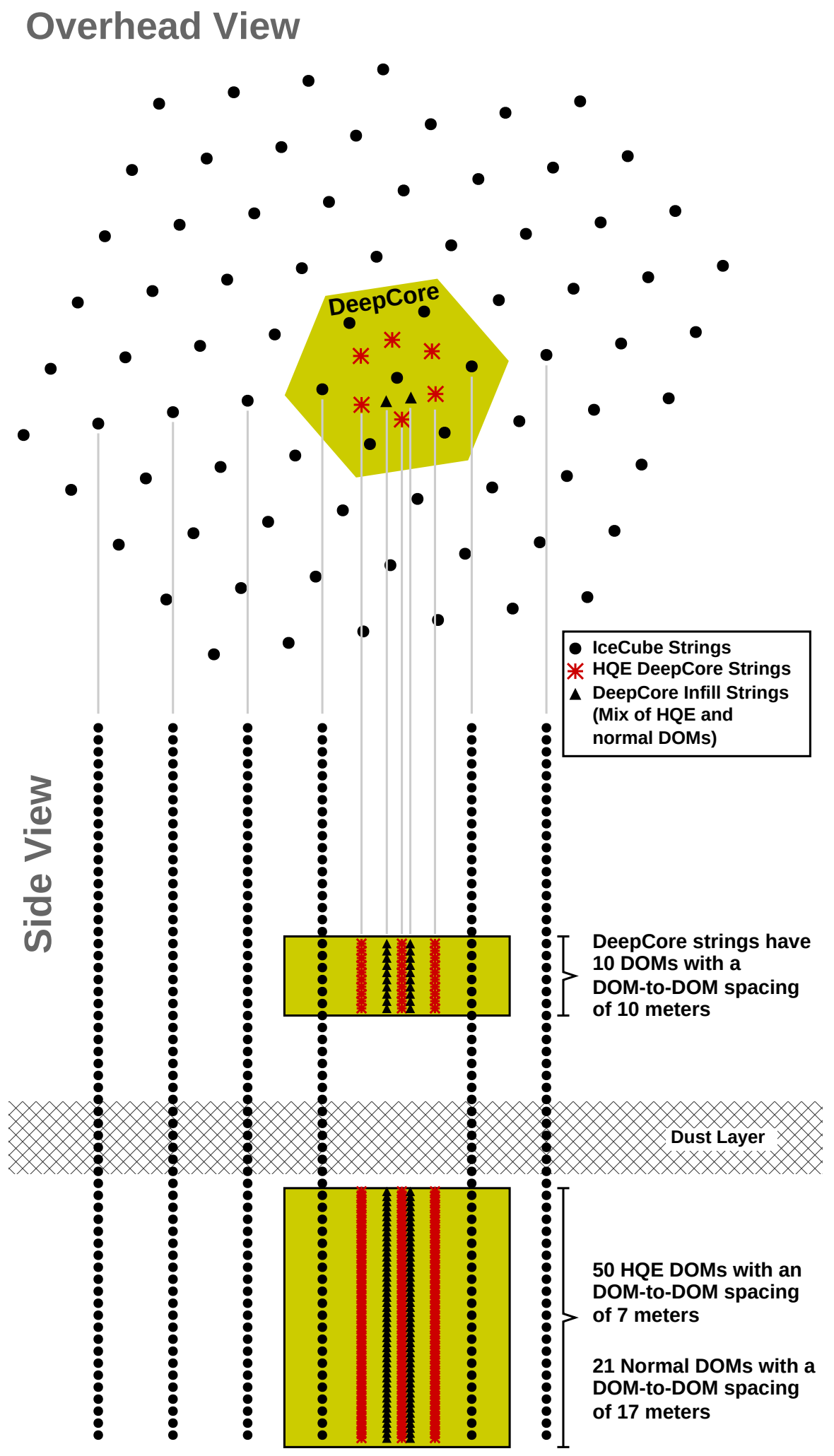

Figure 1: Overhead and side views of the IceCube DeepCore detector. The shaded hexagon in the overhead view shows the area covered by the DeepCore sub-array. On the side view, the hatched region shows the dust layer, and shaded boxes indicate the location of the DeepCore DOMs, 10 in each string above the dust layer, and 50 in each string below the dust layer.

## 2.1 $\nu_\mu$ Disappearance

The earliest atmospheric neutrino oscillation evidence can be traced back to the zenith angle dependence of the double ratio measurement at few GeV energies in Super-Kamiokande [7]. IceCube DeepCore, with its approximately 13 MT fiducial volume, is capable of making atmospheric neutrino oscillation measurements above 10 GeV, an energy region that has not been well explored by previous atmospheric neutrino oscillation experiments. From Fig. 4, a significant deficit in the neutrino flux at 25 GeV is expected from the $\nu_\mu \rightarrow \nu_\mu$ survival probability. In Fig. 5 this disappearance signature is shown for one year of simulated DeepCore data. The disappearance signal assumes that the path length is the diameter of the Earth, and therefore the ideal neutrino sample should contain only up-going neutrino-induced muons. An intrinsic difficulty for all experiments in identifying perfectly up-going neutrino-induced muon tracks is that the average opening angle, defined as the angle between the final state lepton direction and the incoming neutrino direction, increases with decreasing energy. The uncertainty in the opening angle can be approximated as $\Delta\Phi \simeq 30^\circ \times \sqrt{(1GeV)/E_{\nu_\mu}}$. This effect will smear the oscillation signature in the neutrino flux at lower energies. However, as shown in Fig. 5, DeepCore simulations indicate the potential to measure oscillations even if "up-going" tracks are defined to include measured directions over the wide range of values, $-1.0 < \cos(\theta) < -0.6$ [4].

## 2.2 $\nu_\tau$ Appearance

The OPERA neutrino detector, located at the underground Gran Sasso Laboratory (LNGS), was designed for direct observation of $\nu_\mu \rightarrow \nu_\tau$ appearance. OPERA announced a first tau lepton candidate from $\nu_\mu$ oscillation, and continues operation to achieve a statistically significant observation [8, 9]. DeepCore is currently acquiring data and will collect the world's largest inclusive sample of $\nu_\tau$. From Fig. 4, the region where $\nu_\mu$ flux reaches its minimum is the same region where $\nu_\tau$ flux shows its corresponding maximum from $\nu_\mu \rightarrow \nu_\tau$ oscillation. $\nu_\tau$ that interact in DeepCore will produce an electromagnetic or hadronic shower, or "cascade". Events with short tracks which are beyond DeepCore's ability to separate from cascades, are an irreducible background to cascade-like events. Therefore, cascade-like events include: 1) neutral current (NC) events from all three neutrino flavors (e, $\mu$, $\tau$), 2) $\nu_\mu$ charged current (CC) events with short tracks ($< \sim$ O (10) m), 3) $\nu_\tau$ charged current events with $\tau$-leptons decaying into electrons or hadrons, 4) $\nu_\tau$ charged current events with $\tau$-leptons decaying into muons whose track length is less than O (10) m. DeepCore should detect an excess of cascade-like events due to oscillation compared to the number of cascade events expected without oscillation. DeepCore may also be able to detect a distortion in the energy spectrum of cascade events due to $\nu_\tau$ appearance. The simulated excess of cascade-like events above $\sim$ 25 GeV in DeepCore is shown in Fig. 6. The deficit of oscillated cascade-like events compared to unoscillated below $\sim$ 25 GeV is due to the rapid oscillation in the $\nu_\mu$ and $\nu_\tau$ survival probabilities shown in Fig. 4. Within the rapid oscillation regime, a large fraction of $\nu_\mu$ charged current (CC) events oscillate into $\nu_\tau$ events. The produced tau lepton from those $\nu_\tau$ CC events always decay to at least one neutrino, reducing the visible energy and possibly resulting in an event below the detector energy threshold. However, without oscillation, these low energy $\nu_\mu$ CC events would be classified as cascade-like due to their short tracks. This significant deficit between the oscillated and unoscillated cascade-like event rates near the detector threshold may offer an opportunity to measure $\nu_\mu$ disappearance in the cascade channel. The feasibility of measuring the excess of cascade-like events above 25 GeV, as well as systematic effects due to the ice properties, DOM efficiency and other factors, are under study.

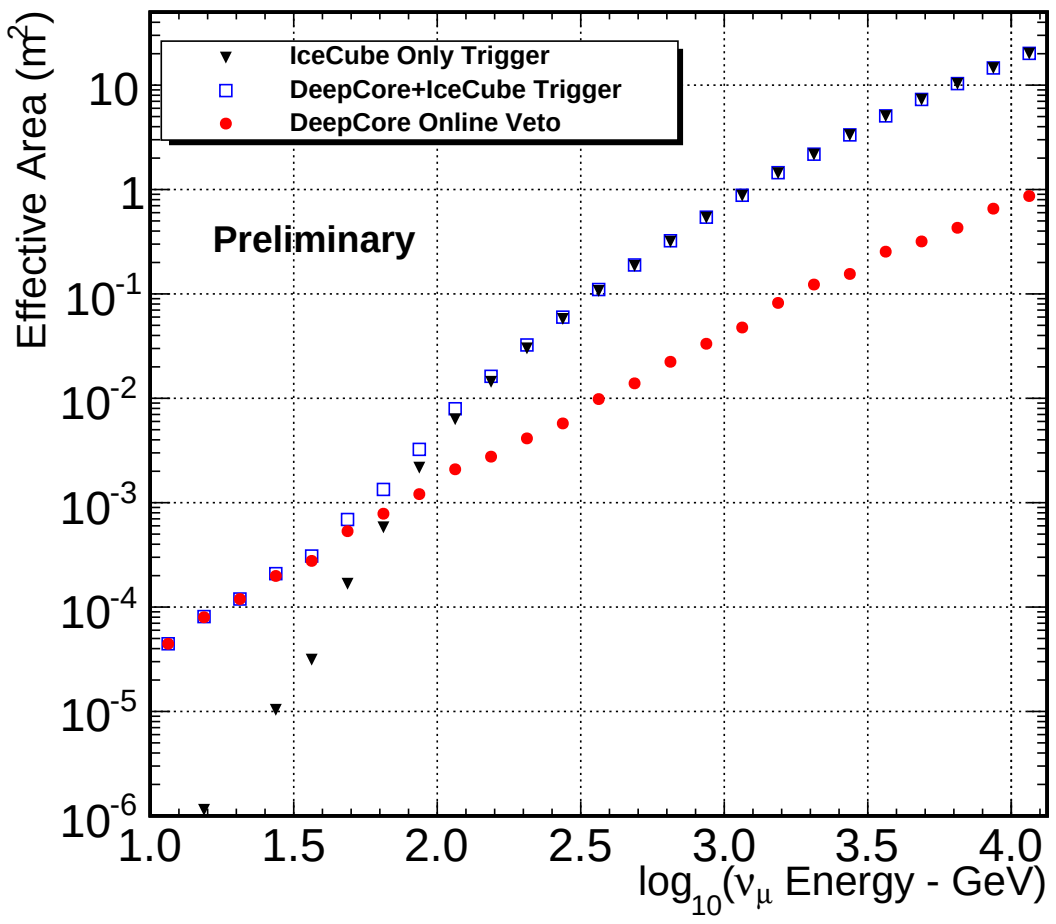

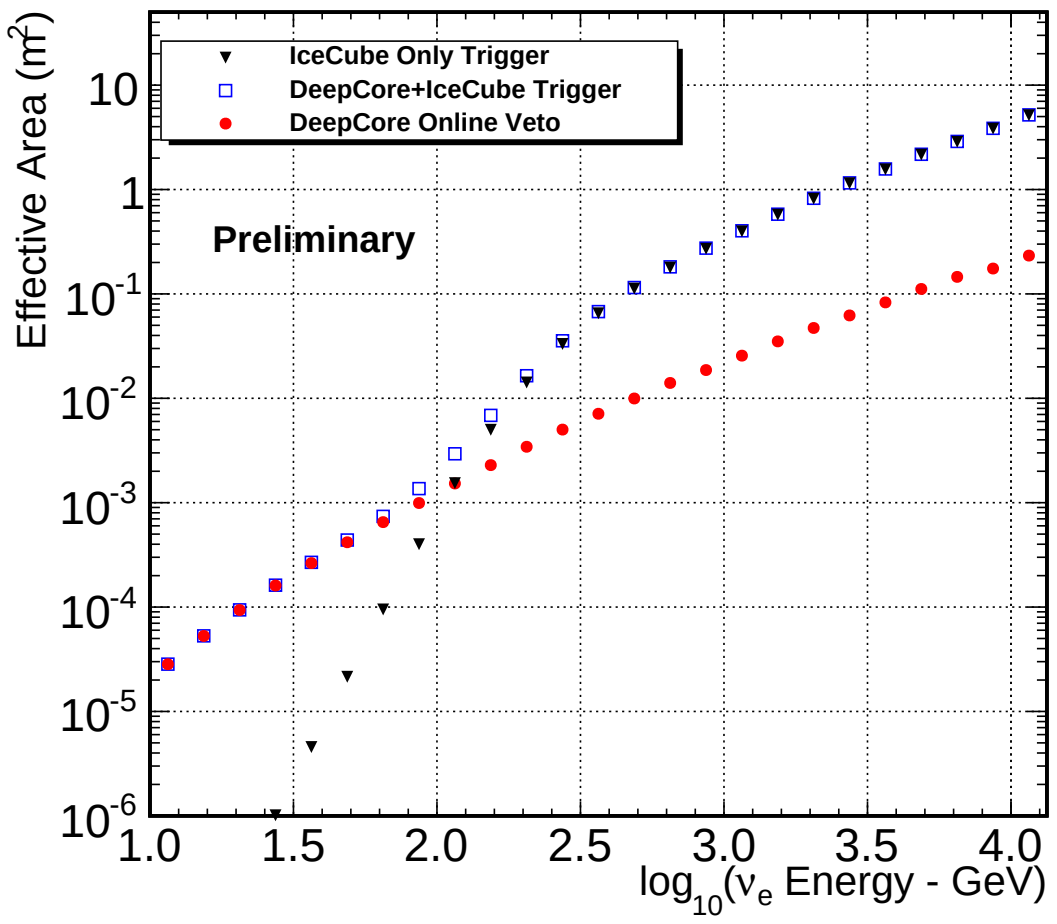

Figure 2: Effective areas for muon neutrinos (left) and electron neutrinos (right). Triangles: IceCube standard strings only, trigger level. Squares: IceCube including DeepCore strings, trigger level. Circles: IceCube DeepCore after applying the online veto.

## 2.3 Reconstruction Algorithms

Several reconstruction algorithms are being developed specifically for low-energy analysis in DeepCore. With shorter muon track lengths, low-energy events include a higher proportion of "starting" and fully contained muon events as opposed to through-going muons. The oscillation analysis also depends on separating track-like $\nu_\mu$ charged current events from cascade-like all-flavor neutral current and $\nu_e$ and $\nu_\tau$ charged current events. Fully-contained-muon reconstruction algorithms calculate the length of the track from the reconstructed beginning and end points of the track. Fig. 7 shows the difference between reconstructed and true track length from simulated data reconstructed with one such algorithm. The starting muon event signature consists of a cascade associated with the charged-current muon neutrino interaction and a track associated with the resulting muon. Reconstructions of such events therefore include the contributions of the cascade and the track. An algorithm under development for seeding these reconstructions and thereby separating tracks from pure cascades is based on a Fermat Surface [10].

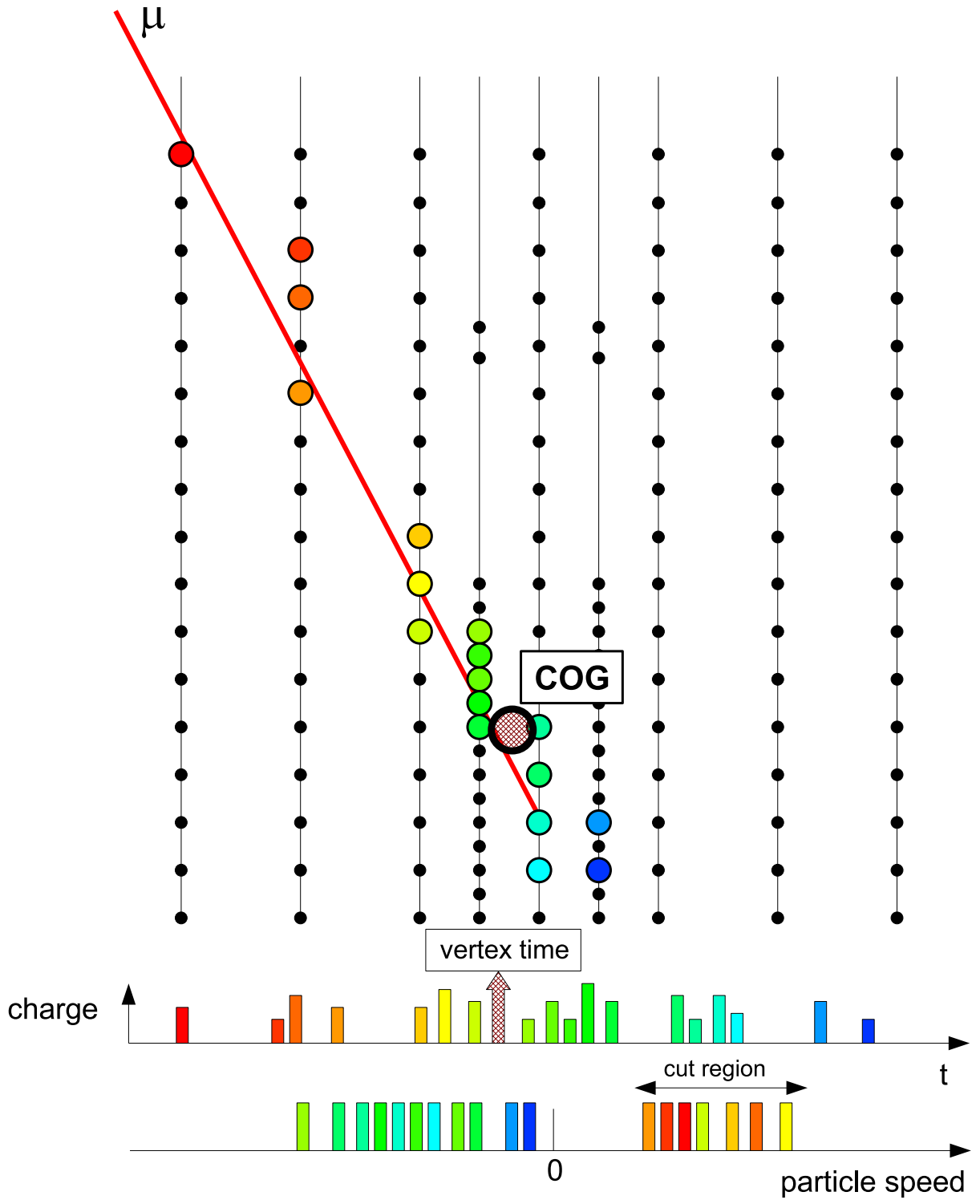

Figure 3: Scheme of veto algorithm based on simulation. Upper part demonstrates a down-going muon hitting the DOMs with hit sequence varying from earliest in red to latest in blue. The big black circle exhibits the COG of the hits in DeepCore. Bottom part includes the vertex time and particle speed per hit. The "cut region" illustrates a cut based on the particle speeds which are consistent with down-going muons.

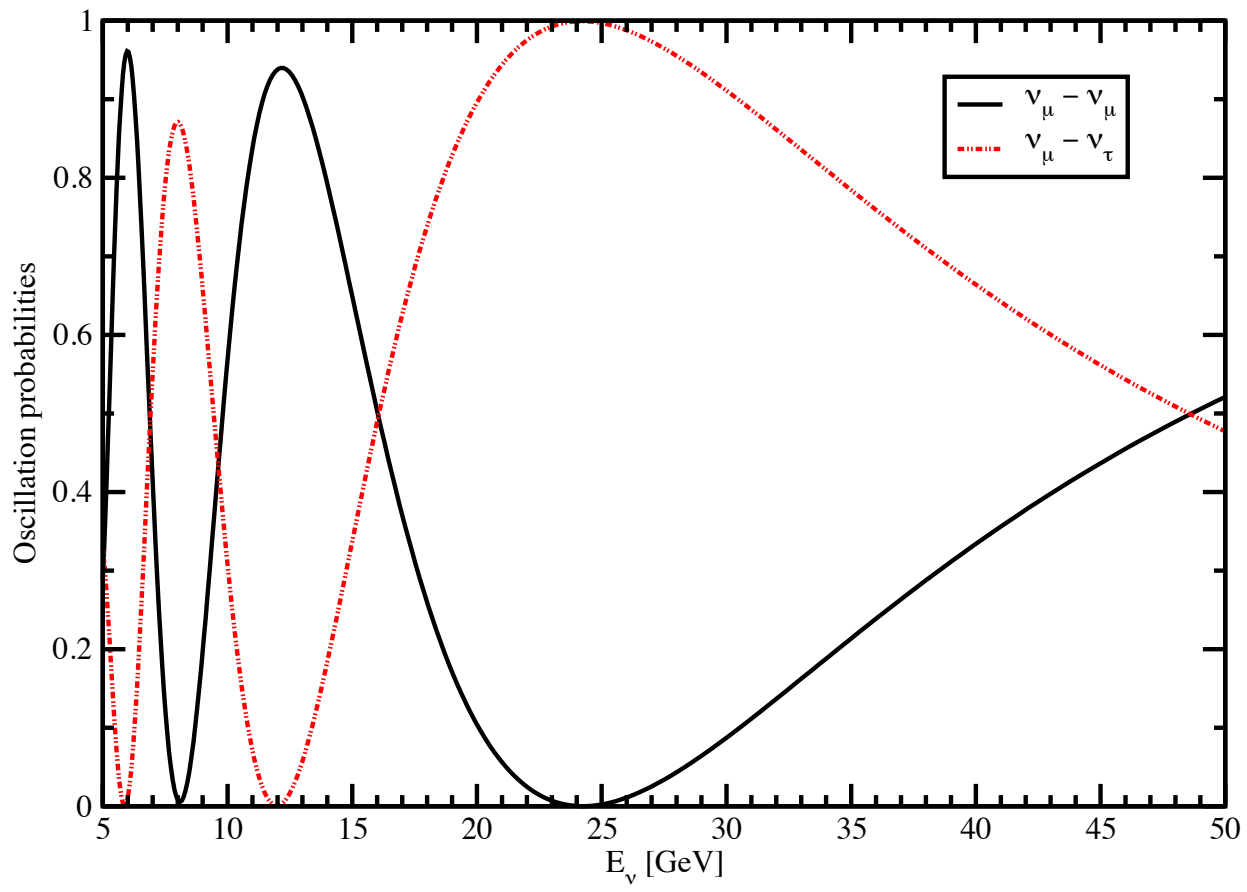

Figure 4: Oscillation probabilities for $\nu_\mu$ with $sin^2 2\theta_{13} = 0.1$, $\Delta m^2_{31} = 2.5 \times 10^{-3} eV^2$, $L = 12757$ km [5].

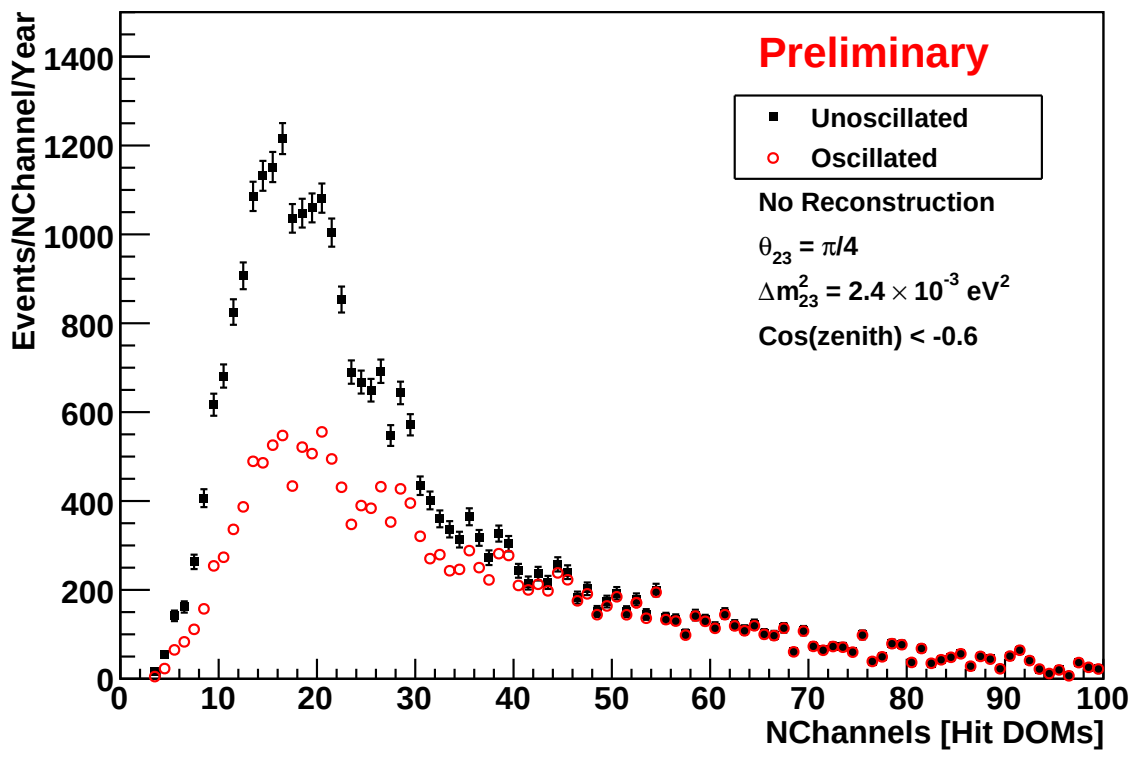

Figure 5: Simulated $\nu_\mu$ disappearance with one year Deep-Core data. Filled squares: Distribution of number of hit DOMs (NChannel) assuming no oscillation. Empty circles: Distribution of number of hit DOMs (NChannel) with oscillation.

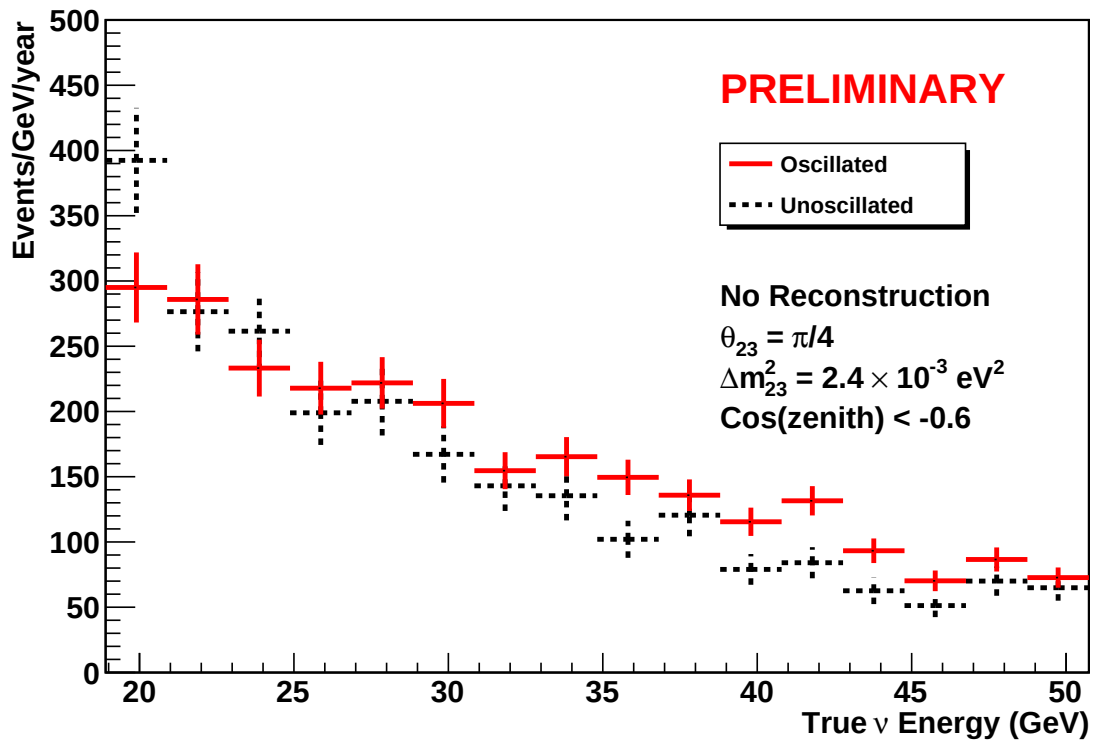

Figure 6: Simulated $\nu_\tau$ appearance with one year Deep-Core data. Dotted lines: Distribution of true neutrino energy assuming no oscillation. Solid lines: Distribution of true neutrino energy with oscillation.

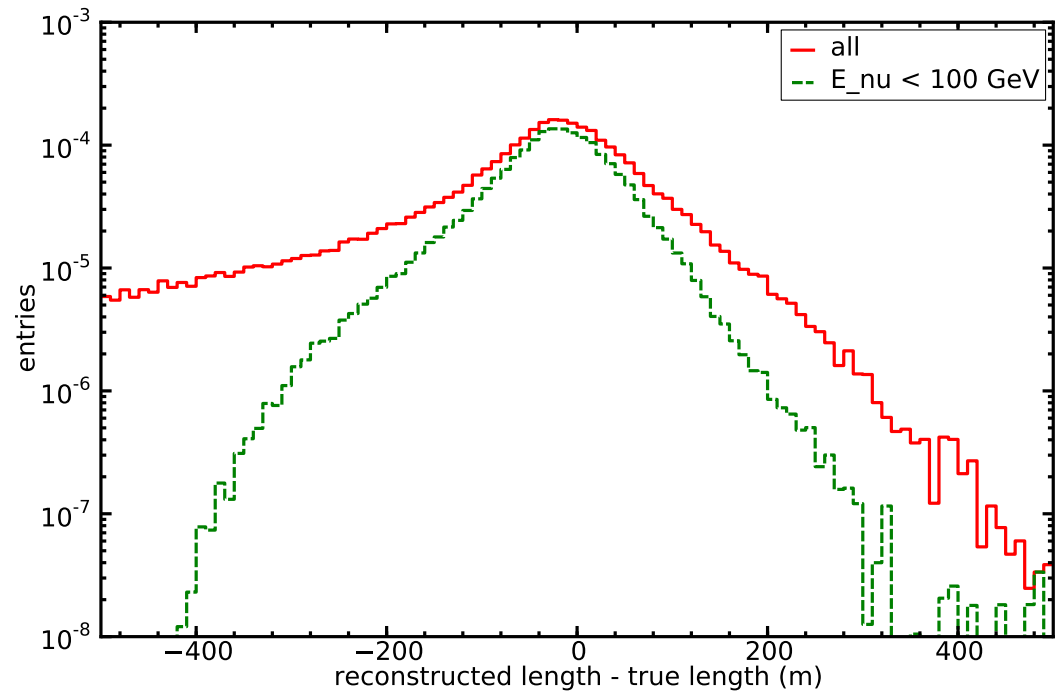

Figure 7: Difference between reconstructed track length and Monte Carlo simulation true track length.

## 3 Conclusions

Simulations suggest that IceCube DeepCore may have the capability to measure atmospheric neutrino oscillations in an energy range which complements existing accelerator measurements. DeepCore construction is complete and the detector is already collecting data. New reconstruction algorithms suited to low energy measurements will enable IceCube to fully exploit the physics capabilities of Deep-Core.

## References

[1] P. B. Price, K. Woschnagg, and D. Chirkin, Geophys. Res. Lett. **27**, 2129-2132 (2000)

[2] A. Achterberg et al. [IceCube Collaboration], Astroparticle Physics, **Volume 26**, Issue 3, October 2006, Pages 155-173

[3] R. Abbasi et al., Nuclear Instruments and Methods **A618** (2010) 139-152, 1-21 June 2010

[4] Darren Grant, D. Jason Koskinen, and Carsten Rott [IceCube Collaboration], Proceedings of the $31^{st}$ ICRC, LODZ, POLAND, 2009

[5] O. Mena, I. Mocioiu and S. Razzaque, Phys. Rev. D **78**, 093003 (2008)

[6] Olaf Schulz, Sebastian Euler, and Darren Grant [Ice-Cube Collaboration], Proceedings of the $31^{st}$ ICRC, LODZ, POLAND, 2009

[7] Y. Fukuda et al. [Super-Kamiokande Collaboration], Phys. Rev. Lett. **81**, 1562 (1998)

[8] R. Acquafredda et al. [OPERA Collaboration], New J. Phys. **8**, 303 (2006)

[9] N. Agafonova et al. [OPERA Collaboration], Phys. Lett. **B691** 138-145 (2010)

[10] J. G. Learned, [arXiv 0902.4009 (2009)]

# Supernova detection with IceCube and beyond

THE ICECUBE COLLABORATION[1]
[1] *See special section in these proceedings*

**Abstract:** In its current configuration, IceCube is a formidable detector for supernovae. It can detect subtle features in the temporal development of MeV neutrinos from the core collapse of nearby massive stars. For a supernova at the galactic center, its sensitivity matches that of a background-free megaton-scale supernova search experiment and triggers on supernovae with about 200, 20, and 6 standard deviations at the galactic center (10 kpc), the galactic edge (30 kpc), and the Large Magellanic Cloud (50 kpc). Signal significances are reduced due to the noise floor and correlations between background hits. In this paper we discuss ways to improve the signal over background ratio with an improved data acquisition. We also discuss methods to track the average neutrino energy by multiple hit detection from individual interacting neutrinos. The latter relies on the ability to reject coincident hits from atmospheric muons.

**Corresponding authors:** V. Baum[1], L. Demirörs[2], L. Köpke[1], M. Ribordy[2] (*mathieu.ribordy@epfl.ch*) , M. Salathe[2], L. Schulte[1]
[1] *Institute of Physics, University of Mainz, Staudinger Weg 7, D-55099 Mainz, Germany*
[2] *Laboratory for High Energy Physics, École Polytechnique Fédérale, CH-1015 Lausanne, Switzerland*

## 1 Introduction

It was recognized early by [1] and [2] that neutrino telescopes offer the possibility to monitor our Galaxy for supernovae. IceCube is uniquely suited for this measurement due to its location and $1\,\mathrm{km}^3$ size. The noise rates in IceCubes photomultiplier tubes average around $540\,\mathrm{Hz}$ since they are surrounded by inert and cold ice with depth dependent temperatures ranging from $-43\,^\circ\mathrm{C}$ to $-20\,^\circ\mathrm{C}$. At depths between $(1450-2450)\,\mathrm{m}$ they are partly shielded from cosmic rays. Cherenkov light induced by neutrino interactions will increase the count rate of all light sensors above their average value. Although this increase in individual light sensor is not statistically significant, the effect will be clearly seen once the rise is considered collectively over many sensors.

The 5160 photomultipliers are installed in modules called digital optical modules (DOMs) and arranged in two configurations: IceCube, with $17\,\mathrm{m}$ ($125\,\mathrm{m}$) vertical spacing between DOMs (horizontal spacing between strings), and DeepCore with $7\,\mathrm{m}$ ($72\,\mathrm{m}$) spacings and equipped with high quantum efficiency DOMs, where $\epsilon_{\mathrm{DeepCore}} = 1.35\epsilon_{\mathrm{IceCube}}$ [3, 4]. With absorption lengths exceeding $100\,\mathrm{m}$, photons travel long distances in the ice such that each DOM effectively monitors several hundred cubic meters of ice. The inverse beta process $\bar{\nu}_e + p \to e^+ + n$ dominates supernova neutrino interactions with $\mathcal{O}(10\,\mathrm{MeV})$ energy in ice or water, leading to positron tracks of about $0.55\,\mathrm{cm}\cdot E_\nu/\mathrm{MeV}$ length for $E_\nu \leq 10\,\mathrm{MeV}$. Considering the approximate $E_\nu^2$ dependence of the cross section, the light yield per neutrino roughly scales with $E_\nu^3$. The detection principle was demonstrated with the AMANDA experiment, IceCubes predecessor [5]. Since 2009, IceCube has been sending real-time datagrams to the Supernova Early Warning System (SNEWS) [6] when detecting supernova candidate events.

Currently, the supernovae search algorithms are based on count rates of individual DOMs stored in $1.67\,\mathrm{ms}$ time bins. We plan to introduce an improved data acquisition system that will allow to store all IceCube hits for supernova candidates. We discuss below some of the improvements that we expect to be achieved using this additional information. In addition, the collaboration is discussing future extensions of the detector that would also improve the supernova detection capacity.

## 2 Current performance

IceCube is the most precise detector for analyzing the neutrino lightcurve of close supernovae. A paper, discussing the detector and physics performance, is close to being published. Figure 1 shows the expected significance for the detection of a supernovae as function of distance (left) and presents the expected rate distribution for the Lawrence-

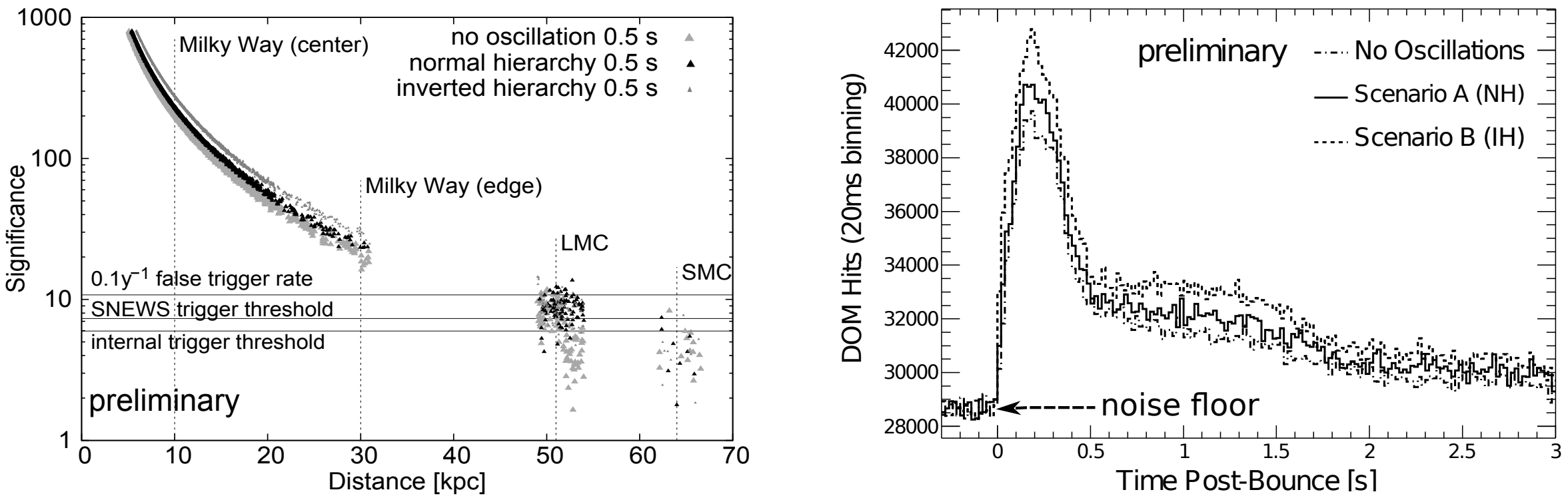

Figure 1: Left: Expected significance versus distance assuming the Lawrence-Livermore model [9] for three oscillation scenarios. The significances are increased by neutrino oscillations in the star by typically 40% in case of an inverted hierarchy. The Magellanic Clouds as well as the center and the edge of the Milky Way and various trigger thresholds are marked. For the Milky Way, the supernova progenitor distribution follows the prediction from [7], for the Magellanic Clouds it is assumed to be uniform. Right: Expected rate distribution at $10\,\mathrm{kpc}$ supernova distance assuming normal and inverse hierarchies.

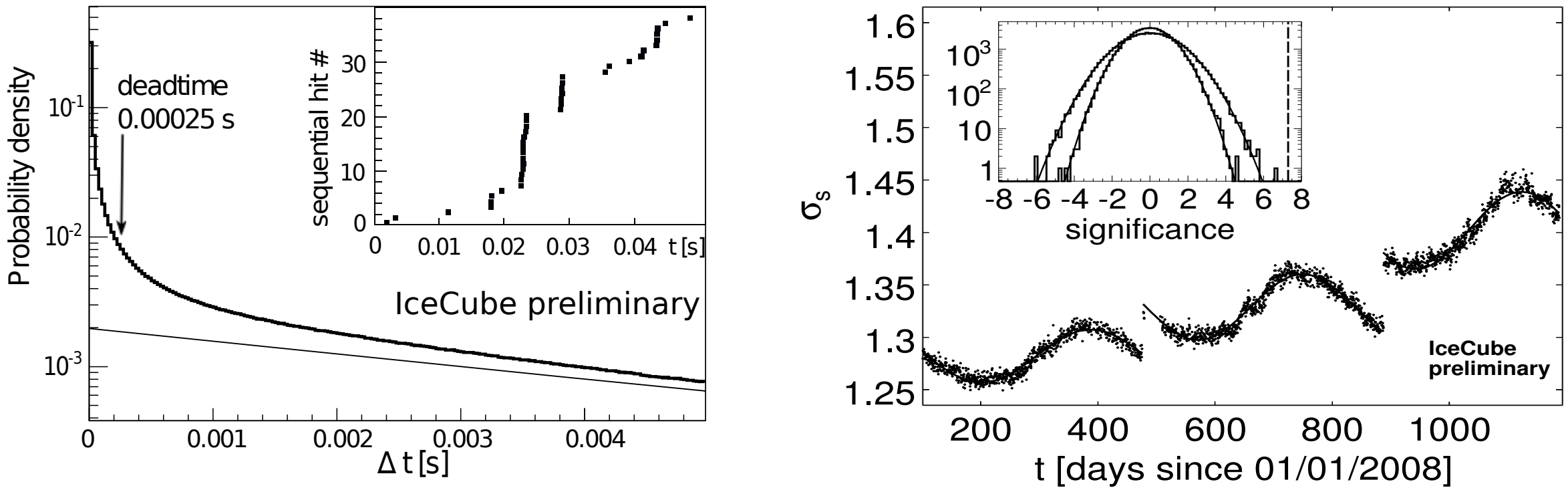

Figure 2: Left: Measured probability density distribution of time differences between pulses for noise (bold line) and the expectation for a Poissonian process fitted in the range $15\,\mathrm{ms} < \Delta T < 50\,\mathrm{ms}$ (thin line). The excess is due to bursts of correlated hits, as indicated by the $50\,\mathrm{ms}$ long snapshot of hit times shown in the inset. Right: Measured width of the significance distribution as function of time during IceCube construction with 40 (left), 59 (middle) and 79 (right) deployed strings. The inset shows the significance distribution before (wide distribution) and after (narrow distribution) suppression of hits due to atmospheric muons (79 strings). The current trigger threshold for SNEWS alarms is indicated by a dashed line.

Livermore model [9] (right). The rate distribution demonstrates the excellent resolution of details in the neutrino light curve. This includes the possibility to distinguish the neutrino hierarchy, provided that the astrophysical model is well known and $\sin^2 \Theta_{13} > 1\,°$[7]. The present noise floor is indicated in Fig. 1 (right) which leads to a fast deterioration of the signal significance particularly at larger distances. In addition, the expected signal significance in IceCube is somewhat reduced due to two types of correlations between pulses that introduce supra-Poissonian fluctuations. The first correlation involves a single photomultiplier tube. It comes about because a radioactive decay in the pressure sphere can produce a burst of photons lasting several $\mu$s. The second correlation arises from the cosmic ray muon background; a single cosmic ray shower can produce a bundle of muons which is seen by hundreds of DOMs. The observed time difference between noise hits deviates from an exponential distribution expected for a Poissonian process (see Fig. 2, left). The inset shows a hit sequence from a single DOM, clearly indicating the bursting behavior. A significant fraction of these bursts can be rejected by an artificial non-paralyzing deadtime, currently adjusted to $\tau = 250\,\mu$s, which decreases the average optical module noise rate to 285 Hz, while keeping $\approx 87\%/(1 + r_{\mathrm{SN}} \cdot \tau)$ of supernova induced hits with rate $r_{\mathrm{SN}}$.

Due to remaining correlated pulses from radioactive decays and atmospheric muons, the measured sample standard deviation in data taken with 79 strings is $\approx 1.3$ and $\approx 1.7$ times larger than the Poissonian expectation for $2\,\mathrm{ms}$ and

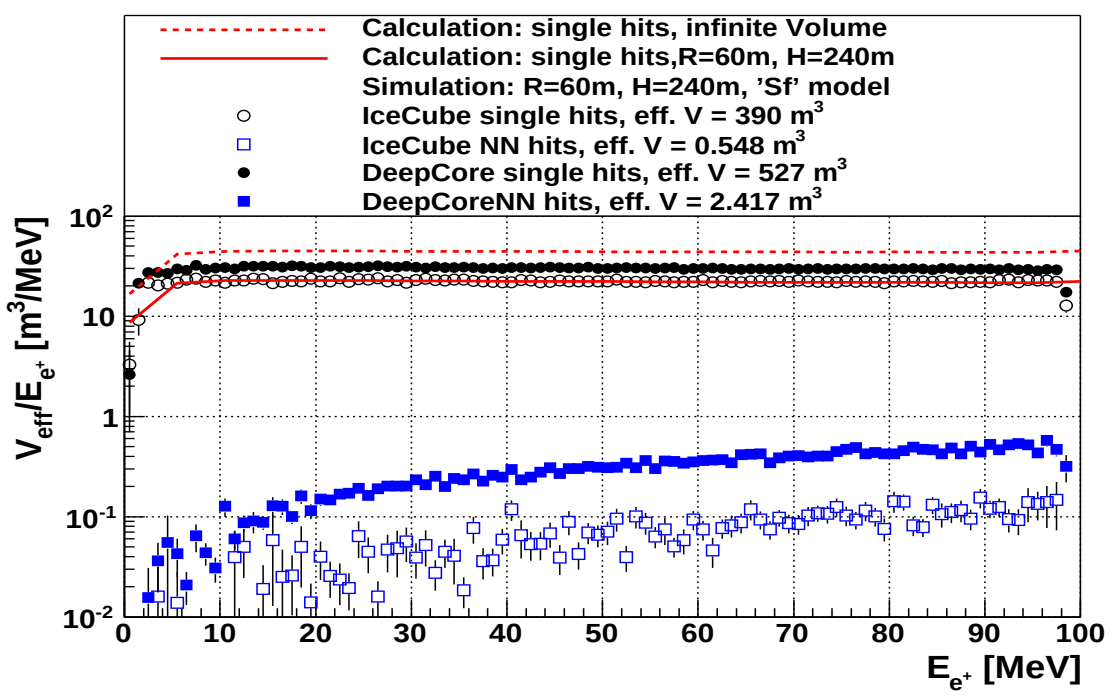

Figure 3: Effective volume per optical module and positron energy versus positron energy. Due to the chosen generation volume, the single hits effective volumes (circles) are about a factor of two too small. This is illustrated by the calculations taken from [11] (lines). The nearest neighbor coincidence modes (squares) are shown here without cut on the coincidence time. The effective volumes given in the legend are calculate according to the ”Sf” model from [12] with an integration time of 4 s.

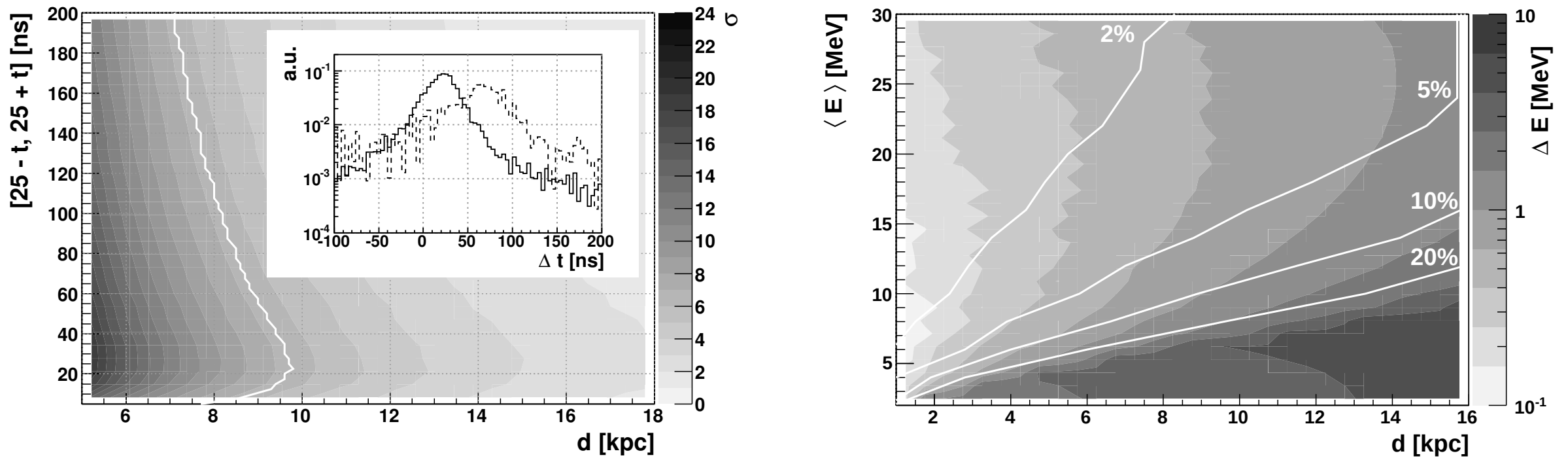

Figure 4: Left: Significance as a function of distance $d$ and the allowed time window $[25\,\mathrm{ns} - t, 25\,\mathrm{ns} + t]$ between hits in two neighboring DOMs 7 m apart (DeepCore). The time difference distribution is shown in the inset in solid for DeepCore and dashed for IceCube. The maximum distance of about 10 kpc is reached for $t = 25\,\mathrm{ns}$ for a significance threshold of 6.7 (white line). Right: Energy resolution $\Delta E$ as a function of distance $d$ and average energy of the emission spectrum $\langle E \rangle$. The white line show several $\Delta E/E$ contours.

500 ms time bins, respectively. For the analysis, we define the significance $\xi = \Delta\mu/\sigma_{\Delta\mu}$ , where $\Delta\mu$ is the most likely collective rate deviation of all optical module noise rates from their running average. $\sigma_{\Delta\mu}$ is the corresponding deviation calculated from the data, thus accounting for non-Poissonian behavior in the dark rates. The significance should be centered at zero with unit width. Fig. 2 (right plot) shows that this is not the case; instead the width changes with the season and increases with the size of the instrumented detector. This behavior is linked to the seasonal change of the muon flux. While atmospheric muons contribute to the count rates of individual DOMs by only 3%, these hits are correlated across the detector, thus broadening the significance distribution and giving rise to a seasonal dependence of the trigger rate. At present, it is possible to subtract roughly half of the hits introduced by atmospheric muons from the total noise rate offline, as the number of coincident hits in neighboring DOMs is recorded for all triggered events. The width of the significance distribution then decreases to about 1.06, close to the expectation (see inset of Fig. 2 (right plot)).

A data acquisition that records all hits in case of a supernova trigger will permit further improvements. The time resolution on the onset of the burst will no longer be restricted by the 1.67 ms time bins in which the rates are recorded, hits associated to triggered atmospheric muons can be fully rejected, and more sophisticated methods to minimize correlated pulse bursts, e.g. by eliminating the initial hits of the bursts while keeping photomultiplier related afterpulses, can be applied. However, as the significance improves only with $1/\sqrt{N_{\mathrm{background}}}$, a much more drastic reduction of background is required in order to improve the detection significance at the edge of our galaxy, to track the average neutrino energy and maybe even provide some directional capability. This can only be achieved by detecting more than one Cherenkov photon from an interaction and applying a coincidence condition.

## 3 New opportunities from coincidence rates

The study in this section is motivated by an analytical framework that explores the potential of coincident hit modes [11]. Here, we investigate the ”nearest neighbor coincidence hits” mode, with a hybrid GEANT-4/toy Monte Carlo simulation. We chose this mode from other possible multi-hit modes such as multiple hits in one DOM or coincident hits between any DOMs, because it has the best noise suppression potential by requiring a very short time window around the two coincident hits.

Figure 3 shows the positron effective volume per module and positron energy for the two detector configurations and two detection modes: single hits and nearest neighbor coincidence hits. The effective volumes given in the legend are calculated according to the ”Sf” model from [12] with an integration time of $4$ s. Both modes have a very different energy dependence, which makes the ratio $N_{\rm coinc}/N_{\rm single}$ an observable of the average energy $\langle E\rangle$ of the emission spectrum.

The inset in Fig. 4 shows the smallest time difference between hits in neighboring DOMs for both detector configurations. A cut on this time distribution was found by calculating the detection significance as a function of the supernova distance and the time window, as is shown in Fig. 4 for DeepCore. Applying a time cut of $\pm 25$ ns around the most probable time difference of $T_0 = 25$ ns, a maximum distance of $10$ kpc can be reached with the DeepCore DOM separation. A similar cut yields a smaller reach for IceCube due to the larger DOM separation.

To estimate the energy resolution of the energy observable $N_{\rm coinc}/N_{\rm single}$, the ratio was calculated using spectra according to the ”Sf” model with average energies between $2-30$ MeV. In Fig. 4 (right plot), the deviation $\Delta E$ is shown as a function of the supernova distance and the average $\langle E\rangle$. For distance smaller than the trigger threshold ($\approx 10$ kpc), a resolution of around 5% can be achieved for spectra with an average energy of $10-15$ MeV.

The energy resolution depends on the expected noise level for the chosen selection mode. Above, the Poissonian noise levels were scaled up by 1.3, as mentioned in Sec. 2. It is possible that this underestimates the average hit probability for DOMs that were close to atmospheric muons. Preliminary studies show that light from muons can be suppressed by considering only DOMs that were around $300$ m or further away from the reconstructed track position. In the worst case, when a track traverses the whole detector volume vertically, this cut reduces the usable volume by $\approx 30\%$. Alternatives to such a cut are being investigated.

## 4 Possible extensions of IceCube/DeepCore

Discussions on an extension of IceCube/DeepCore have started, which would also improve the supernova detection capability. We used a GEANT-4 simulation to estimate the capabilities of a hypothetic 18 string detector with IceCube DOMs spaced apart a few meters. Fig. 5 shows the effective volumes of single and multiple hits as function of distance between the DOMs. The sizable increase in the active volume of coincidence hits would strongly improve the signal over noise ratio and lead e.g. to a substantial improvement of a supernova detection at the Magellanic cloud.

## 5 Conclusion

As a supernova detector, IceCube already offers an unmatched ability to establish subtle features in the temporal development of the neutrino flux by tracking the overall count rates of its DOMs. The correlated noise background remains one of the big challenges. The new data acquisition will permit the recording of all hits in case of a trigger signal, thus greatly improving the rejection of correlated noise sources such as atmospheric muons. It will also allow to study the influence of the correlated noise on multiple hit detection modes.

One such mode, nearest neighbor coincident hits, was introduced in this paper and shows great potential to extent the existing capabilities of IceCube by measuring the average neutrino energy. This mode also has the potential to be sensitive to the neutrino direction. Other multi-hit modes will be studied next. All these modes would greatly benefit from a very dense sub-array with inter-DOM and inter-string distance of a few meters only, extending their reach to several tens of kilo parsec and possibly beyond.

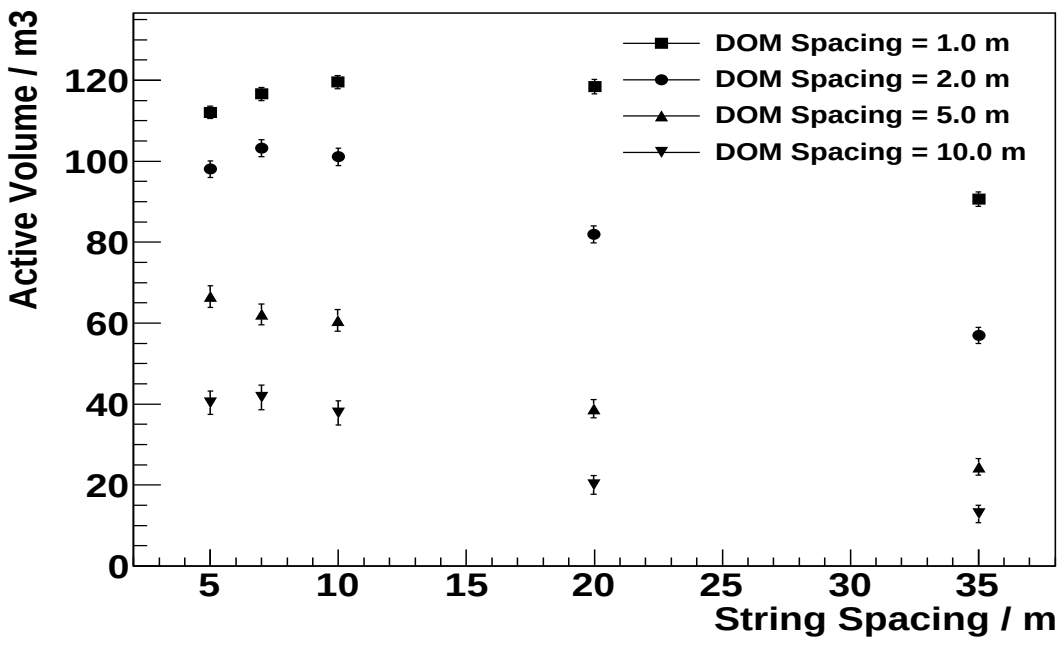

Figure 5: Effective volumes per DOM for multiple hits as function of horizontal and vertical distance between DOMs. The same neutrino spectrum was used as in Fig. 3

## References

[1] C. Pryor, C. E. Roos, M. S. Webster, Astrophys. J. **329**, 335 (1988).

[2] F. Halzen, J. E. Jacobsen, E. Zas, Phys. Rev. **D53**, 7359-7361 (1996).

[3] [IceCube Collab.] J. Ahrens *et al.*, Astropart. Phys. **20**, 507-532 (2004).

[4] [IceCube Collab.] C. Wiebusch, arXiv:0907.2263

[5] J. Ahrens *et al.*, Astropart. Phys. **16**, 345-359 (2002).

[6] P. Antonioli, R. T. Fienberg, F. Fleurot, Y. Fukuda, W. Fulgione, A. Habig, J. Heise, A. B. McDonald *et al.*, New J. Phys. **6**, 114 (2004).

[7] J. N. Bahcall, T. Piran, Astrophys. J. **267**, L77 (1982).

[8] [IceCube Collab.] R. Abbasi *et al.*, in progress

[9] T. Totani, K. Sato, H. E. Dalhed, J. R. Wilson, Astrophys. J. **496**, 216-225 (1998).

[10] A. S. Dighe, M. T. Keil, G. G. Raffelt, arXiv:hep-ph/0303210v3.

[11] M. Ribordy, arXiv:submit/0263088 [astro-ph.IM]

[12] L. Hudepohl, B. Muller, H. -T. Janka, A. Marek, G. G. Raffelt, Phys. Rev. Lett. **104**, 251101 (2010).

# Study of South Pole ice transparency with IceCube flashers

THE ICECUBE COLLABORATION[1]
[1] *See special section in these proceedings*

**Abstract:** The IceCube observatory, 1 km$^3$ in size, is now complete with 86 strings deployed in the antarctic ice. IceCube detects the Cherenkov radiation emitted by charged particles passing through or created in the ice. To realize the full potential of the detector the properties of light propagation in the ice in and around the detector must thus be known to the best achievable accuracy. This report presents a new method of fitting the ice model to a data set of in-situ light source events collected with IceCube. The resulting set of derived ice parameters is presented and a comparison of IceCube data with simulation based on the new model is shown.

**Corresponding author:** Dmitry Chirkin (*dima@icecube.wisc.edu*)
IceCube Reseach Center, University of Wisconsin, Madison, WI 50703, U.S.A.

## 1 Introduction

The properties of photon propagation in a transparent medium can be described in terms of the average distance between successive scatters and the average distance to absorption (local scattering and absorption lengths), as well as the angular distribution of the new direction of a photon relative to old at a given scattering point. These details are used in both the simulation and reconstruction of IceCube data, thus they must be known to the best possible accuracy. This work presents a new, *direct fit* approach to determine these ice properties, which is different from the method described in [4]. A global fit is performed to a set of data with in-situ light sources (see Figure 1) covering all depths of the detector, resulting in a single set of scattering and absorption parameters of ice, which describes these data best. Figure 2 shows examples of experimental data used for this analysis.

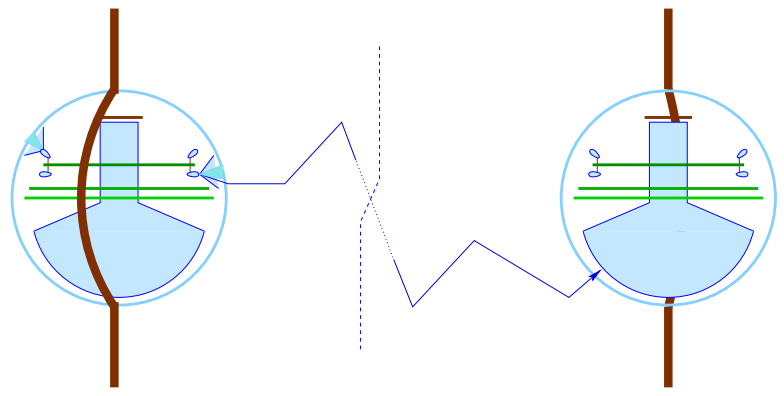

Figure 1: Simplified schematics of the experimental setup: the flashing sensor on the left emits photons, which propagate through ice and are detected by a receiving sensor on the right.

## 2 Flasher dataset

In 2008, IceCube consisted of 40 strings as shown in Figure 3, each equipped with 60 equally spaced optical sensors, or digital optical modules (DOMs). Each of the DOMs consists of a 10” diameter photomultiplier tube (PMT) [2] and several electronics boards enclosed in a glass container [3]. One of the boards is the ”flasher board”, which has 6 horizontal and 6 tilted LEDs, each capable of emitting $\sim 7.5 \cdot 10^9$ photons at $\sim 405 \pm 5$ nm in a 62 ns-wide pulse.

The PMT output signal is digitized into ”waveforms” using the faster, ATWD, and slower, fADC, sampling chips [1]. The ATWD is configured to collect 128 samples with 3.3 ns sampling rate, and the fADC records 256 samples with 25 ns sampling rate. The DOMs transmit time-stamped digitized PMT signal waveforms to computers at the surface.

In a series of several special-purpose runs, IceCube took data with each of 60 DOMs on string 63 flashing in a sequence. For each of the flashing DOMs at least $250$ flasher events were collected and used in this analysis. All 6 horizontal LEDs were used simultaneously at maximum brightness and pulse width settings, creating a pattern of light around string 63 that is approximately azimuthally symmetric.

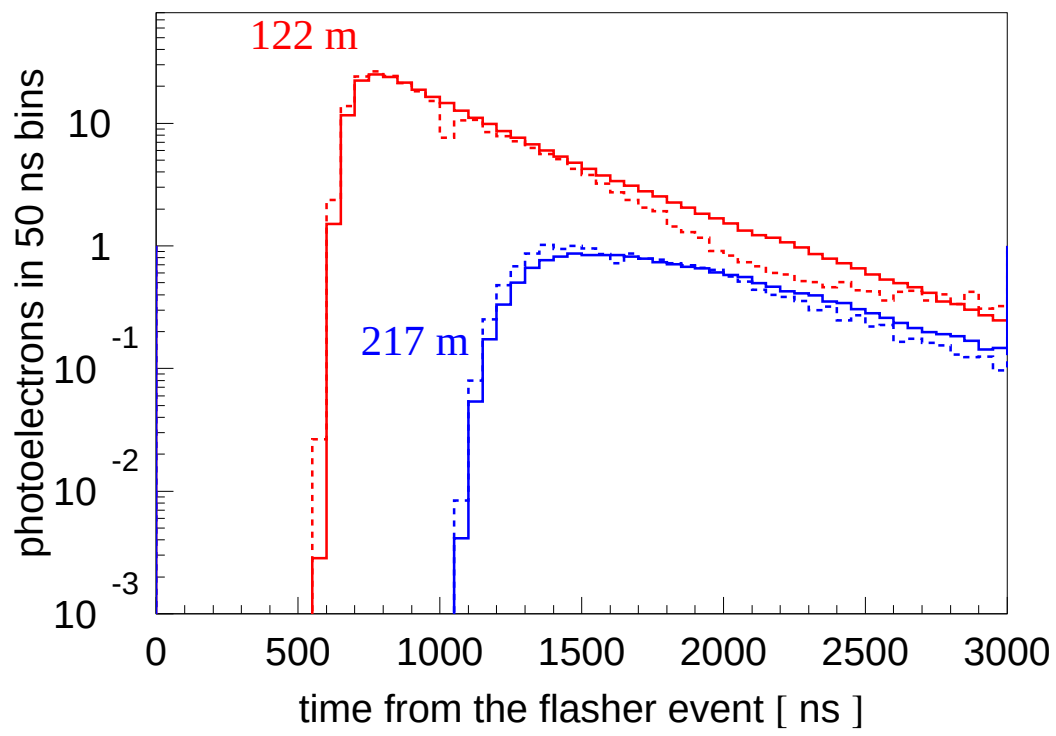

Figure 2: Example photon arrival time distributions at a sensor on one of the nearest strings (122 m away), and on one of the next-to-nearest strings (217 m away). Dashed lines show data and solid lines show simulation based on the model of this work (with best fit parameters). The goal of this work is to find the best-fit ice parameters, which describe these distributions as observed in data simultaneously for all pairs of emitters and receivers.

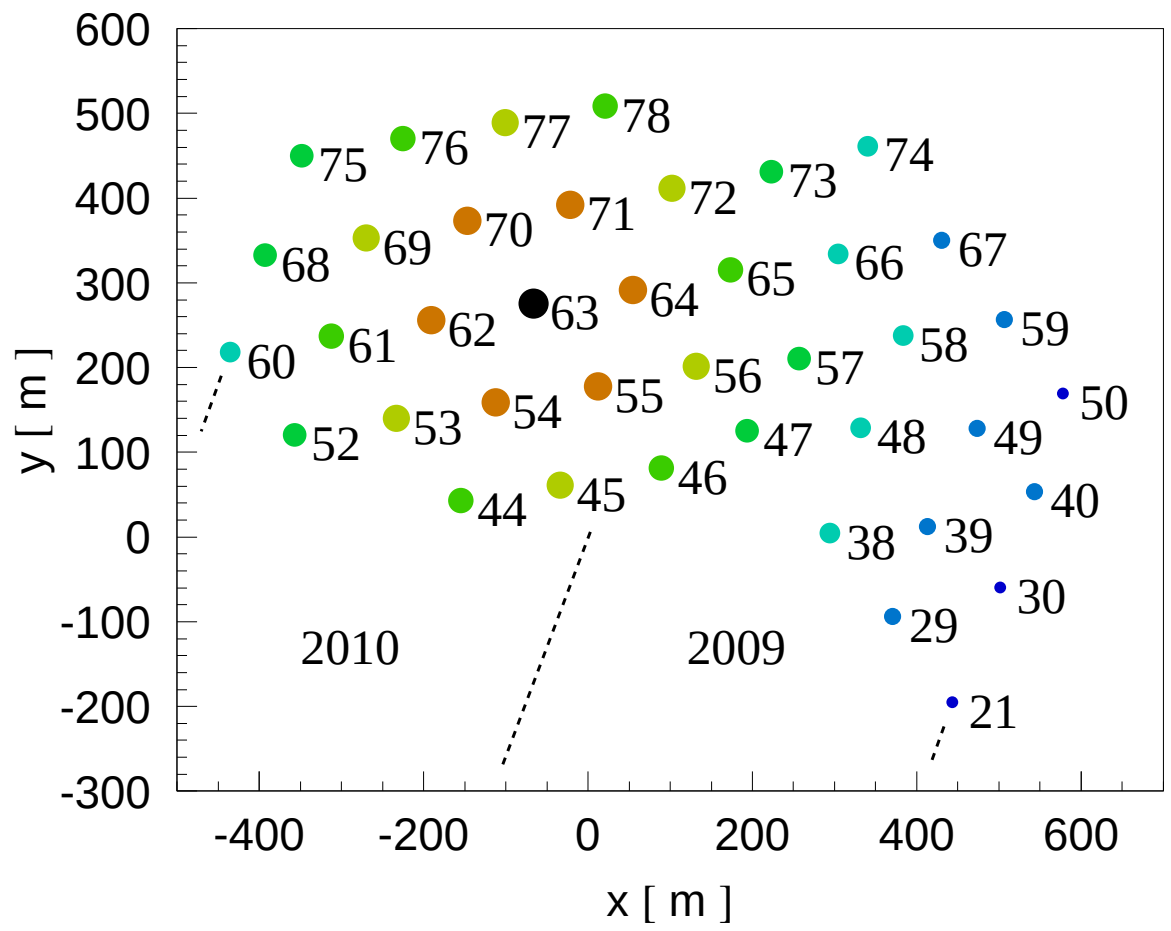

Figure 3: IceCube 40-string configuration as operated in 2008. String 63 (of DOMs that were used as flashers) is shown in black. IceCube parts installed in the following years (2009, 2010 as shown in the figure) lie in regions indicated approximately with dashed lines.

The pulses corresponding to the arriving photons were extracted from the digitized waveforms and binned in 25 ns bins, from 0 to 5000 ns from the start of the flasher pulse (extracted from the special-purpose ATWD channel of the flashing DOM). To reduce the contribution from saturated DOMs (most of which were on string 63 near the flashing DOM) [2] the photon data collected on string 63 was not used in the fit.

## 3 Ice parametrization

The ice is described by a table of parameters $b_e(405)$, $a(405)$, related to scattering and absorption at a wavelength of 405 nm at different depths. The width of the vertical ice layers (10 m) was chosen to be as small as possible while maintaining at least one receiving DOM in each layer. Coincidentally it is the same as the value chosen in [4].

The geometrical scattering coefficient $b$ determines the average distance between successive scatters (as $1/b$). It is often more convenient to quote the effective scattering coefficient, $b_e = b \cdot (1 - \langle\cos\theta\rangle)$, where $\theta$ is the deflection angle at each scatter, $\langle\rangle$ denote the expectation value. The absorption coefficient $a$ determines the average distance traveled by photon before it is absorbed (as $1/a$).

## 4 Simulation

The detector response to flashing each of the 60 DOMs on string 63 needs to be simulated very quickly, so that simulations based on many different sets of coefficients $b_e(405)$ and $a(405)$ could be compared to the data.

A program called PPC (photon propagation code [7]), was written for this purpose. It propagates photons through ice described by a selected set of parameters $b_e(405)$ and $a(405)$ until they hit a DOM or get absorbed. No special weighting scheme was employed, except that the DOMs were scaled up in size (a factor 5 to 16, depending on the required timing precision), and the number of emitted photons was scaled down by a corresponding factor ($5^2 - 16^2$).

The probability distribution $f(\theta)$ of the photon scattering angle $\theta$ is modeled by a linear combination of two functions commonly used to approximate scattering on impurities:

$$f(\theta) = (1 - f_{\mathsf{SL}}) \cdot \mathsf{HG} + f_{\mathsf{SL}} \cdot \mathsf{SL},$$

where HG is the Henyey-Greenstein function [4]:

$$p(\cos\theta) = \frac{1}{2}\frac{1 - g^2}{[1 + g^2 - 2g \cdot \cos\theta]^{3/2}}, \quad g = \langle\cos\theta\rangle,$$

and SL is the simplified Liu scattering function [8]:

$$p(\cos\theta) \sim (1 + \cos\theta)^\alpha, \quad \text{with} \quad \alpha = \frac{2g}{1 - g}.$$

$f_{\mathsf{SL}}$ determines the relative fraction of the two scattering functions and it determines the overall shape. Figure 4 compares these two functions with the prediction of the Mie theory with dust concentrations and radii distributions taken as described in [4]. The distributions of photon arrival time are substantially affected by the "shape" parameter $f_{\mathsf{SL}}$ (as shown in Figure 5). $f_{\mathsf{SL}}$ is also a global free parameter in the fitting procedure.

The value of $g = 0.9$ was used in this work (cf. $g = 0.8$ in [4]). Higher values (as high as $\sim 0.94$ [4, 6]) are predicted by the Mie scattering theory, however, these result in slower simulation, while yielding values of the effective scattering $b_e$ and absorption $a$ coefficients that change by less than 3% as determined in [4], which could also be concluded from Figure 5.

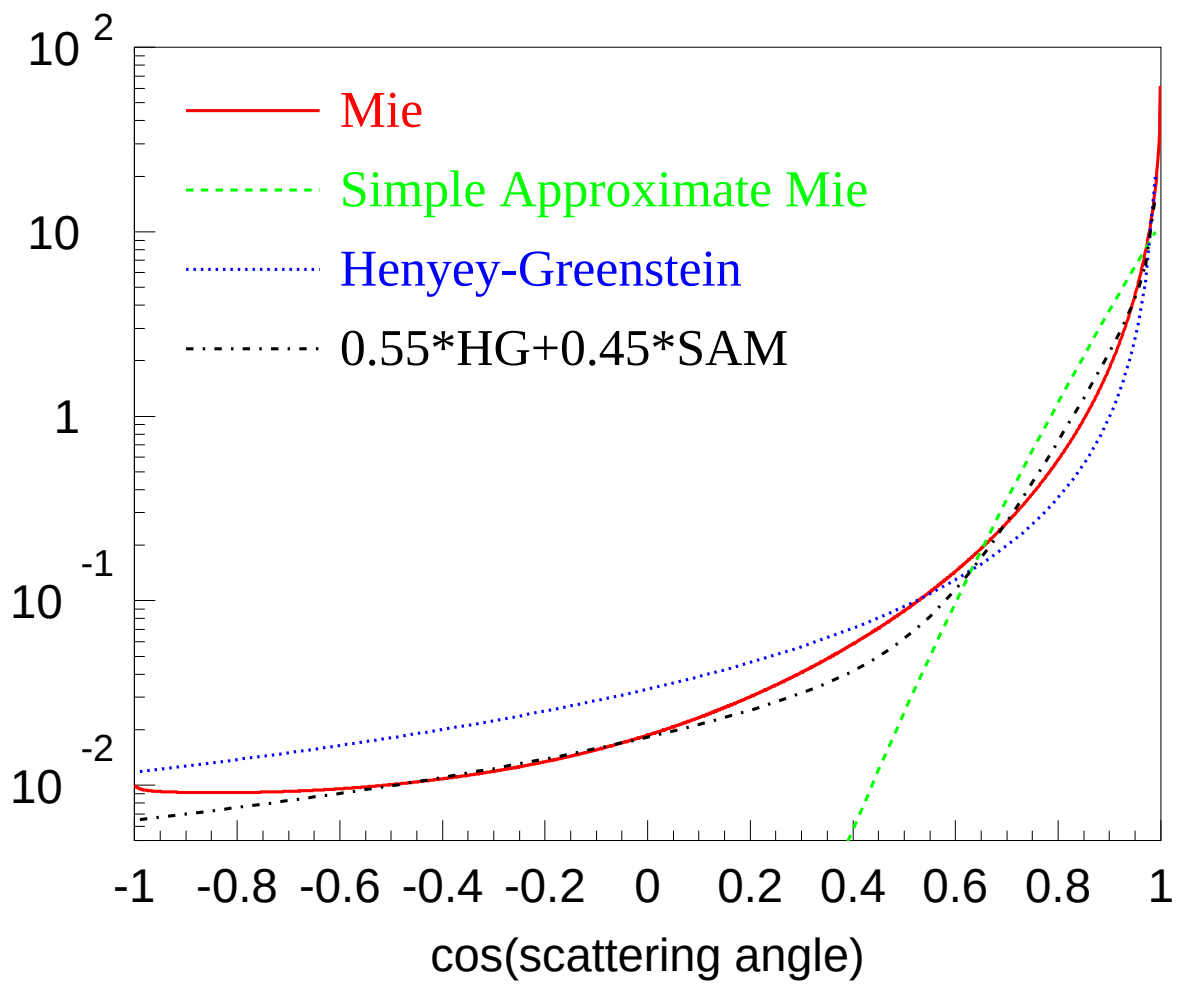

Figure 4: Comparison of the Mie scattering profiles calculated at several depths of the South Pole ice with the Henyey-Greenstein (HG) [4] and simplified Liu (SL) [8] scattering functions, all with the same $g = 0.943$.

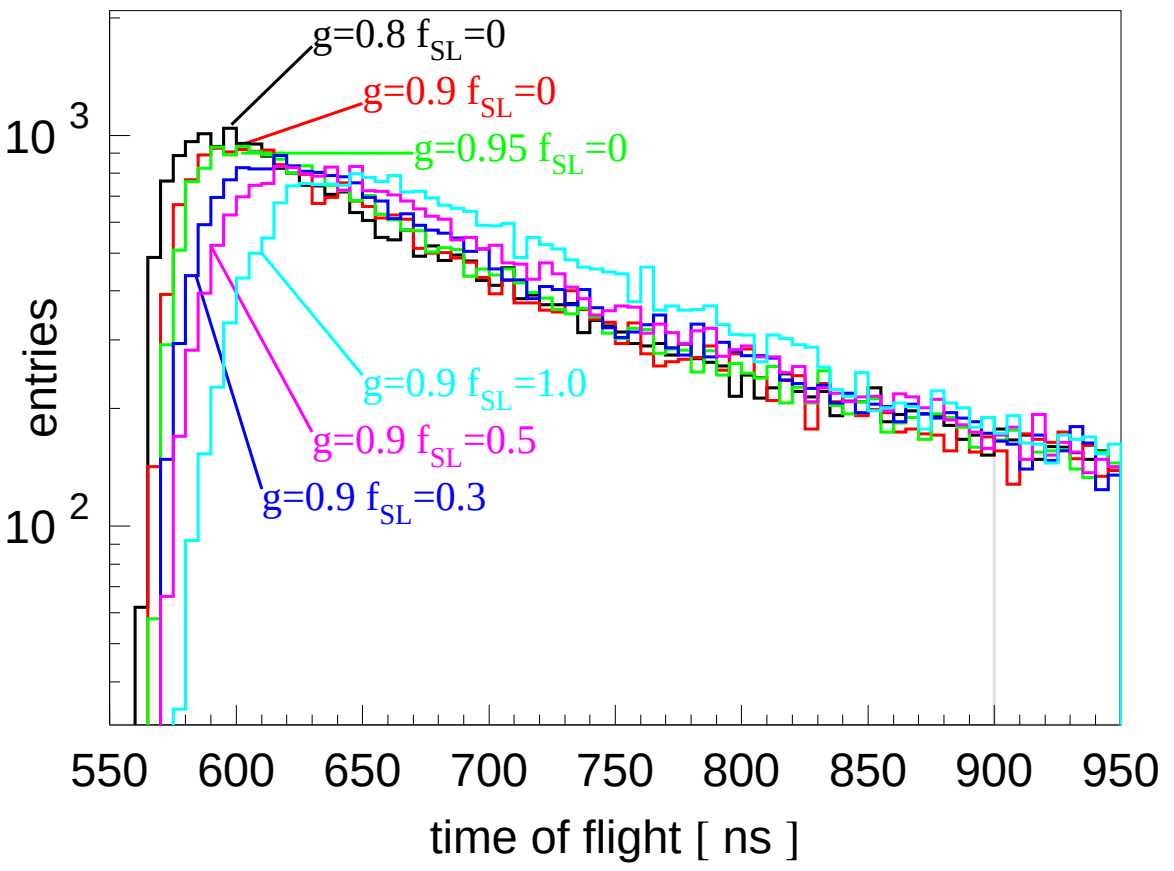

Figure 5: Photon arriving time distributions at a DOM 125 m away from the flasher, simulated for several values of $g = \langle\cos\theta\rangle$ and $f_{\mathrm{SL}}$. The difference in peak position simulated with $g = 0.8$ and $g = 0.9$ is of the same order ($\sim$ 10 ns) as that between sets simulated with different values of the shape parameter $f_{\mathrm{SL}}$.

## 5 Fitting the flasher data

Data from all pairs of emitter-receiver DOMs (located in the same or different ice layers, altogether $\sim$ 38700 pairs) contributed to the fit of $\sim$ 200 ice parameters (scattering

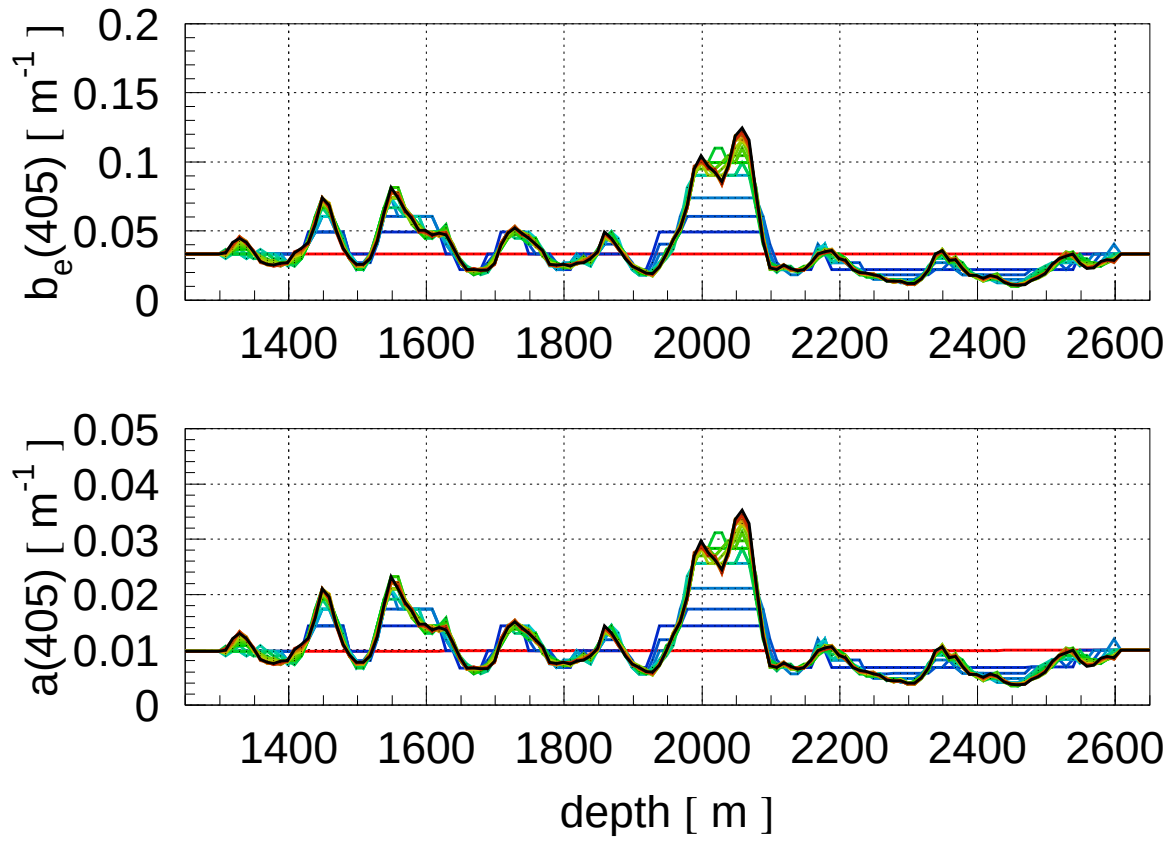

Figure 6: Values of $b_e(405)$ and $a(405)$ vs. depth for subsequent steps of the minimizer. The "converged" black curve shows fitted values after the last of the 20 steps of the minimizer.

and absorption in 10 m layers at detector depths of 1450 to 2450 m).

The photon counts $d(t_i)$ and $s(t_i)$ observed in time bins in $n_d$ data and $n_s$ simulated flasher events are compared to each other using a likelihood function

$$\mathcal{L} = \prod \frac{(\mu_s n_s)^s}{s!} e^{-\mu_s n_s} \cdot \frac{(\mu_d n_d)^d}{d!} e^{-\mu_d n_d} \cdot \frac{1}{\sqrt{2\pi}\sigma} \exp \frac{-(\log \mu_d - \log \mu_s)^2}{2\sigma^2} \cdot R.$$

The product is over all emitters and all time bins of receivers. $\mu_d(t_i)$ and $\mu_s(t_i)$ are the expected values of photon counts per event in data and simulation, and are determined by maximizing $\mathcal{L}$ with respect to these. The first two terms in the product are the Poisson probabilities, and the third term describes the systematic uncertainties inherent in the simulation. The last term $R$ represents regularization constraints of the solution values with depth and with each other.

Starting with the homogeneous ice described with $b_e(405) = 0.042\ \mathrm{m}^{-1}$ and $a(405) = 8.0\ \mathrm{km}^{-1}$ (average of [4] at detector depths) the maximum of $\mathcal{L}$ is found in $\sim$ 20 steps. At each iteration step the values of $b_e(405)$ and $a(405)$ are varied in consecutive ice layers, one layer at a time. Five flashing DOMs closest to the layer, which properties are varied, are used to estimate the variation of the $\mathcal{L}$. Figure 6 shows ice properties after each of 20 steps of the minimizer. The general agreement of the model and data is good as shown in Figure 2.

## 6 Dust logger data

Several dust loggers [5] were used during the deployment of seven of the IceCube strings to result in a survey of the structure of ice dust layers with extreme detail (with the effective resolution of $\sim$ 2 millimeters). These were then matched up across the detector to result in a *tilt map* of the

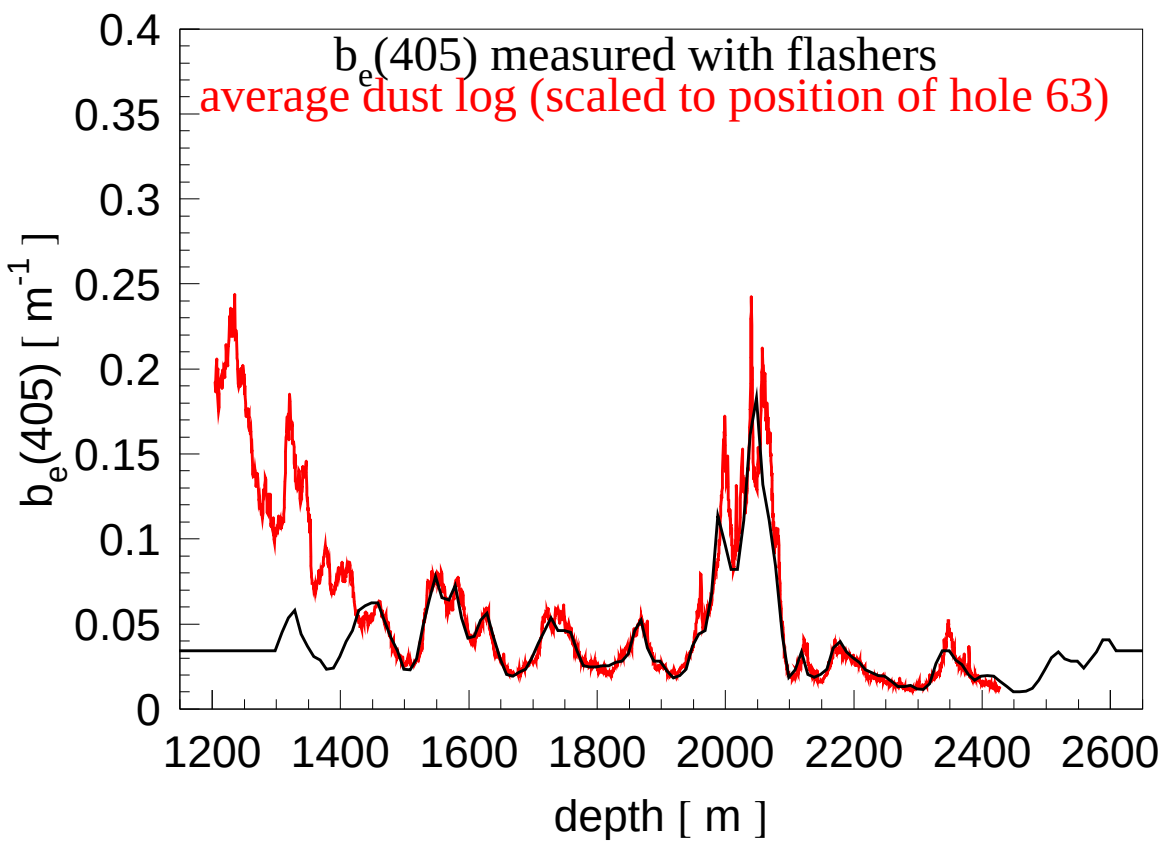

Figure 7: Comparison of the average dust log with the effective scattering coefficient $b_e(405)$ measured with the flasher data.

South Pole ice, as well as a high-detail *average dust log*, a record of a quantity proportional to the dust concentration vs. depth. Additionally, the EDML (East Dronning Maud Land, see [5]) ice core data was used to extend the dust record to below the lowest dust-logger-acquired point.

The correlation between the effective scattering coefficient measured with the IceCube flasher data and the average dust log (scaled to the location of string 63) is excellent, as shown in Figure 7. Within the depth range 1450 m - 2450 m instrumented with DOMs all major features match, have the right rise and falloff behavior, and are of the same magnitude. Some minor features are washed out in the flasher measurement.

Having established the correlation with the average dust log, the EDML-extended version of the log was used to build an initial approximation to the fitting algorithm described in the previous section. This resulted in a solution that is determined by the scaled values of the extended log (instead of by the somewhat arbitrary values of the initial homogeneous ice approximation) in the regions where the flasher fitting method has no resolving power, i.e., above and below the detector.

## 7 Results

The effective scattering and absorption parameters of ice measured in this work are shown in Figure 8 with the $\pm 10\%$ gray band corresponding to $\pm 1\sigma$ uncertainty at most depths. The uncertainty grows beyond the shown band at depths above and below the detector. The value of the scattering function parameter $f_{\mathsf{SL}} = 0.45$ was also determined.

Figure 8 also shows the AHA (Additionally Heterogeneous Absorption) model, which is based on the ice description of [4] extrapolated to cover the range of depths of IceCube and updated with a procedure enhancing the depth structure of the ice layers.

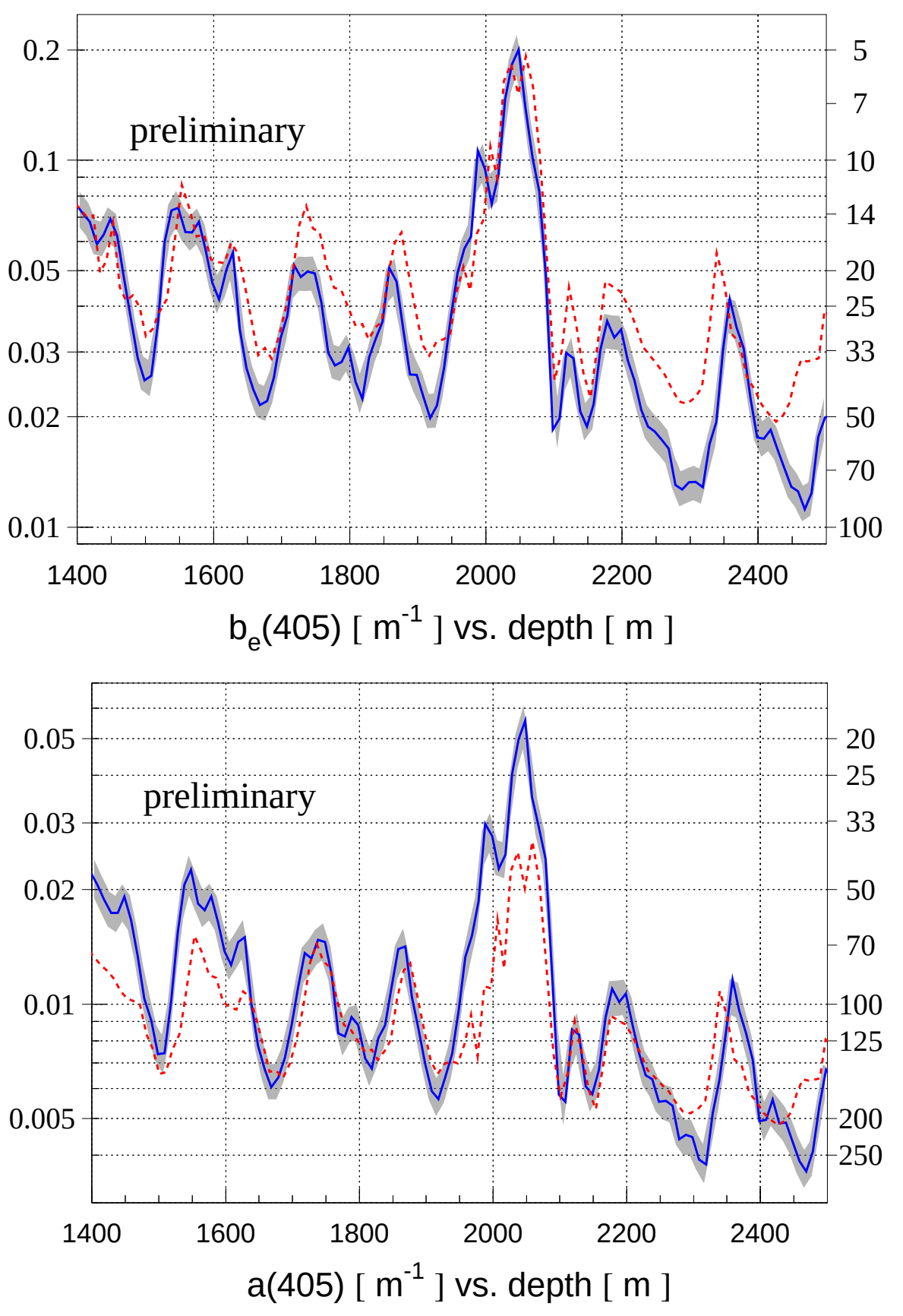

Figure 8: Values of $b_e(405)$ and $a(405)$ vs. depth for converged solution shown with solid lines. The updated model of [4] (AHA) is shown with dashed lines. The uncertainties of the AHA model at the AMANDA depths of $1730 \pm 225$ m are $\sim 5\%$ in $b_e$ and $\sim 14\%$ in $a$. The scale and numbers to the right of each plot indicate the corresponding effective scattering $1/b_e$ and absorption $1/a$ lengths in meters.

## References

[1] R. Abbasi et al., Nucl. Instrum. Methods A, 2009, **601**: 294-316.

[2] R. Abbasi et al., Nucl. Instrum. Methods A, 2010, **618**: 139-152.

[3] A. Achterberg et al., Astropart. Physics, 2006, **26**: 155.

[4] M. Ackermann et al., J. Geophys. Res., 2006, **111**: D13203.

[5] R. C. Bay et al., J. Geophys. Res., 2010, **115**: D14126.

[6] D. Chirkin., 2002, Mie scattering code and plots: http://icecube.wisc.edu/~dima/work/ICESCA.

[7] D. Chirkin., 2010, photon propagation code: http://icecube.wisc.edu/~dima/work/WISC/ppc.

[8] Pingyu Liu., Phys. Med. Biol., 1994, **39**: 1025.